\documentclass{aa}  
\usepackage{graphicx,natbib}
\usepackage{txfonts}
\newcommand{\Msun}{\ensuremath{\mathit{M}_{\odot}}}
\newcommand{\Teff}{\ensuremath{T_{\mathrm{eff}}}}
\newcommand{\Ni}{\ensuremath{ ^{56}\mathrm{Ni}}}

\newcommand{\kin}{\mathrm{kin}}
\newcommand{\esc}{\mathrm{esc}}

\newcommand{\mrunit}{\ensuremath{M_\odot~\mathrm{yr^{-1}}}}
\newcommand{\kmps}{\ensuremath{\mathrm{km~s^{-1}}}}

\begin{document} 

   \authorrunning{Moriya \& Langer}
   \titlerunning{Pulsations of RSG PISN progenitors and their mass loss}

   \title{
Pulsations of red supergiant pair-instability supernova
progenitors leading to extreme mass loss
 }

   \subtitle{}

   \author{Takashi J. Moriya
          \and
	  Norbert Langer
          }

   \institute{
          Argelander Institute for Astronomy, University of Bonn,
          Auf dem H\"ugel 71, D-53121 Bonn, Germany
   \\
          \email{moriyatk@astro.uni-bonn.de}
             }

   \date{Received 11 September 2014 / Accepted 15 October 2014}

\abstract{
Recent stellar evolution models show consistently that
very massive metal-free stars evolve into red
supergiants shortly before they explode. We argue that the
envelopes of these stars,
which will form pair-instability supernovae, become pulsationally
unstable and that this will lead to extreme mass-loss rates 
despite the tiny metal content of the envelopes.
We investigate the pulsational properties of such models and derive
pulsationally induced mass-loss rates, which take the damping effects
of the mass loss on the pulsations selfconsistently into account. 
We find that the pulsations may induce mass-loss rates of 
$\sim 10^{-4}-10^{-2}$ \mrunit\ 
shortly before the explosions, which may create a dense circumstellar
medium.
Our results show that very massive stars with dense circumstellar media
may stem from a wider initial mass range than 
pulsational-pair instability supernovae.
The extreme mass loss will cease when so much of the hydrogen-rich envelope is lost that the
star becomes more compact and stops pulsating.
The helium core of these stars therefore remains unaffected,
and their fate as pair-instability supernovae remains unaltered.
The existence of dense circumstellar media around metal-free pair-instability
supernovae can make them brighter and bluer, and they may be easier 
to detect at high redshifts than previously expected.
We argue that the mass-loss enhancement in pair-instability supernova
progenitors can naturally explain some observational properties of superluminous supernovae:
the energetic explosions of stars within hydrogen-rich dense circumstellar media
with little \Ni\ production and
the lack of a hydrogen-rich envelope in pair-instability supernova
candidates with large \Ni\ production.
}

 

   \keywords{stars: evolution -- stars: massive -- stars: mass-loss -- supernovae: general}

   \maketitle
%

\section{Introduction}\label{introduction}
Pair-instability supernovae (PISNe) are theoretically predicted
explosions of very massive stars that are triggered by the creation
of electron-positron pairs in their cores \citep[e.g.,][]{rakavy1967,barkat1967}.
The pair creation makes very massive stars dynamically unstable and
leads to their collapse.
The collapse triggers explosive
nuclear burning, thereby unbinding the entire star.
PISNe can be very bright because of their huge explosion energy
and large \Ni\ production \citep[e.g.,][]{kasen2011,dessart2013,whalen2013,kozyreva2014}.
PISNe have long remained a theoretical prediction, but several
possible observational candidates have been discovered recently \citep[e.g.,][]{gal-yam2009}.
The predicted characteristic chemical signatures of PISNe
\citep[e.g.,][]{heger2002}
have not yet been observed
in metal-poor stars \citep[e.g.,][]{yong2013},
although a potential candidate metal-poor star showing such a signature
has been reported recently \citep{aoki2014}.

\begin{table*}
\caption{Properties of our stellar models:
Conversion efficiency,
initial mass, final mass,
final hydrogen-rich envelope mass,
final helium core mass,
final radius, final effective temperature,
and final luminosity.
} 
\label{tablestandard}      
\centering          
\begin{tabular}{c c c c c c c c}
\hline\hline       
$\varepsilon$ &$M_\mathrm{ini}$ & $M_\mathrm{fin}$ &
 $M_\mathrm{fin}^\mathrm{env}$ &
$M_\mathrm{fin}^\mathrm{core}$ &
 $R_\mathrm{fin}$ & $T_\mathrm{eff,fin}$ & $L_\mathrm{fin}$ \\
&($M_\odot$) & ($M_\odot$) & ($M_\odot$) &($M_\odot$) &
($10^3 R_\odot$) & ($10^3$ K) & ($10^6 L_\odot$)\\
\hline  
 &150 & 147 & 73 & 74 &3.14&4.63&4.09\\
0$^{a}$&200 & 190 & 87 & 103 &3.78&4.71&6.31\\
 &250 & 236 & 111& 126 &4.16&4.72&7.73\\
\hline
   &150 & 104 & 30 & 74 &2.70&4.99&4.13\\
0.1$^{b}$&200 & 136 & 32 & 104 &3.44&4.94&6.34\\
   &250 & 164 & 40 & 124 &3.81&4.94&7.82\\
\hline  
   &150 & 104 & 31 & 73 &2.75&4.97&4.13\\
0.3$^{b}$&200 & 129 & 25 & 104 &3.52&4.90&6.40\\
   &250 & 170 & 45 & 125 &3.90&4.89&7.78\\
\hline  
\end{tabular}
\tablefoot{
\tablefoottext{a}{Models with radiation-driven wind only.} 
\tablefoottext{b}{Models in which pulsation-driven mass loss with the indicated
value of $\varepsilon$ is considered.} 
}
\end{table*}

Strong mass loss on the main sequence or thereafter may lead to lower core masses,
thereby preventing very massive stars developing the pair instability.
Thus, PISNe are likely to occur where the mass-loss rates can be low
\citep[e.g.,][]{heger2002,umeda2002,langer2007,yoon2012,yusof2013,yoshida2014}, i.e., at low metallicity.
Especially, the first zero-metallicity stars are
said to form very massive stars primarily, possibly producing
PISNe due to the lack of radiation-driven mass loss \citep[e.g.,][]{hirano2014,susa2014}.
Despite their low metallicity, stars that are slightly less massive than
PISN progenitors are able to experience a large amount of mass
loss very late in their evolution \citep[e.g.,][]{woosley2007,ohkubo2009,chatzopoulos2012,chatzopoulos2012b}.
This is because the pair instability occurs in these stars, but it is so weak that
they are only partially disrupted. The successive eruptions of these stars
may produce a dense circumstellar medium (CSM), which 
may explain the high luminosities in some superluminous
SNe (SLSNe, \citealt{woosley2007,whalen2014,chen2014}).
Even though these luminous transients are likely not accompanied by SNe
since the stars eventually collapse to form a black hole,
they are called pulsational pair-instability SNe.

An interesting property of PISN progenitors displayed by stellar
evolution models is that many of them are red supergiants (RSGs)
when they explode \citep[e.g.,][]{langer2007,yoon2012,dessart2013}.
An important property of RSGs that has not been taken into account
in the PISN progenitor modeling is
that hydrogen-rich envelopes of RSGs are pulsationally unstable when their
$L/M$-ratio is high, where
$L$ is the stellar luminosity and $M$ the stellar mass \citep{li1994,heger1997,yoon2010}.
The fundamental mode of the radial pulsations is said to grow in
RSGs due to the $\kappa$ mechanism working at the hydrogen ionization zones
in the RSG envelopes.

Pulsational properties of metal-free very massive stars have been
investigated by \citet{baraffe2001}. While they were found to be
pulsationally unstable on the main sequence, \citet{baraffe2001} conclude from
the tiny growth rates of the pulsations that any effect on the mass-loss
rates would be negligible \citep[see also][]{sonoi2012}.
Their most massive models (300 and 500~\Msun) evolved into RSGs at the end of their evolution,
and while they noted that the models became more violently pulsationally unstable
during that stage, they did not investigate this any further.

\begin{figure}
\centering
\includegraphics[width=\columnwidth]{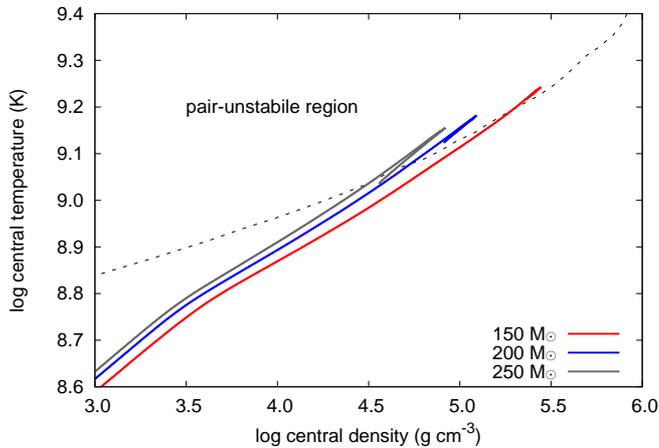}
\caption{
Evolution of central density and temperature of our PISN
 progenitor models. In the `pair-unstable region', the stars become
dynamically unstable.
}
\label{central}
\end{figure}

Various authors suggested that for large enough amplitudes, 
pulsations may induce stellar mass loss
\citep[e.g.,][]{appenzeller1970a,appenzeller1970b,bowen1988,hoefner2003,neilson2008}.
For example, pulsations are known to be an essential driver of the mass loss
in carbon-rich asymptotic giant-branch stars.
The pulsations create high-density regions above the stellar photosphere
in which dust can be formed. 
Thanks to the high opacity of the formed dust, their mass-loss rates are
enhanced significantly \citep[e.g.,][]{hoefner2003,hoefner2008,wood1979}.
The large mass loss induced by pulsations in the RSG stage was said
to influence the final fates of massive stars \citep[e.g.,][]{heger1997,yoon2010,moriya2011,georgy2012}.
Previous studies of RSG pulsations focussed on
core-collapse SN progenitors below 40 \Msun\ \citep{heger1997,yoon2010}.

In this paper, we investigate the pulsational properties of massive
metal-free PISN progenitors ($150-250\ \Msun$) during their RSG stage.
After introducing our methods to follow the stellar evolution in Section~\ref{evolution},
we show that these stars are pulsationally unstable,
discuss their pulsational properties, and derive their
pulsation-induced mass-loss rates in Section~\ref{pulsation}.
The effect of the induced mass loss
on the evolution and explosions of PISN progenitors are discussed
in Section~\ref{consequences}.
We conclude this paper in Section~\ref{conclusions}.

\begin{figure}
\centering
\includegraphics[width=\columnwidth]{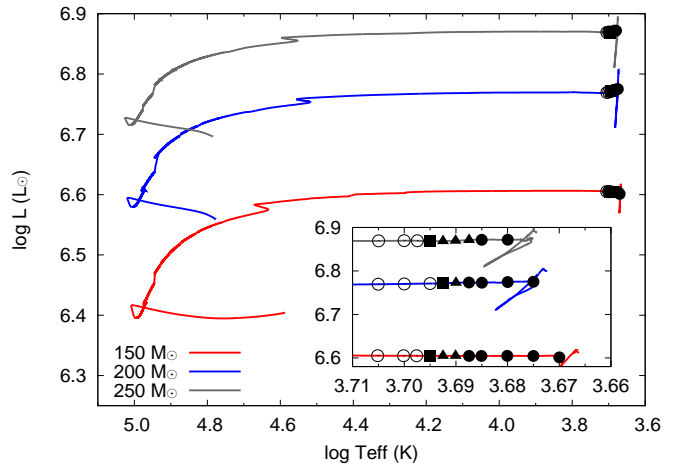}
\caption{
Evolution of our PISN progenitor models in the HR diagram.
Those models marked with the symbols
are used to examin their pulsational properties.
We do not find unstable pulsations in models marked by open circles.
The models indicated with squares have the pulsational pattern A,
those with triangles have the pattern B, and
those with filled squares have the pattern C.
See Fig.~\ref{standard} for the definitions of the patterns.
}
\label{hrd}
\end{figure}

\begin{figure*}
\centering
\includegraphics[width=\columnwidth]{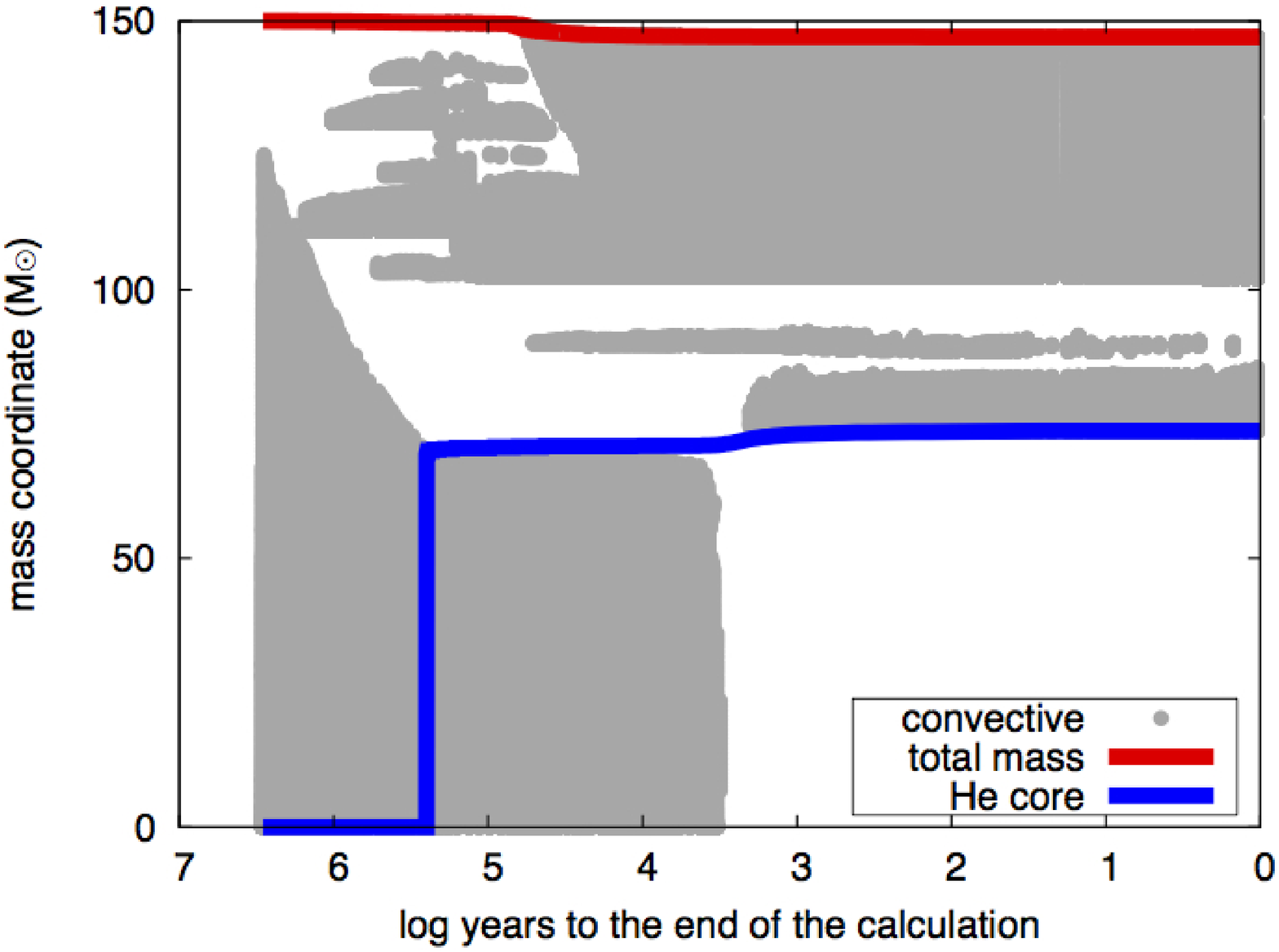}
\includegraphics[width=\columnwidth]{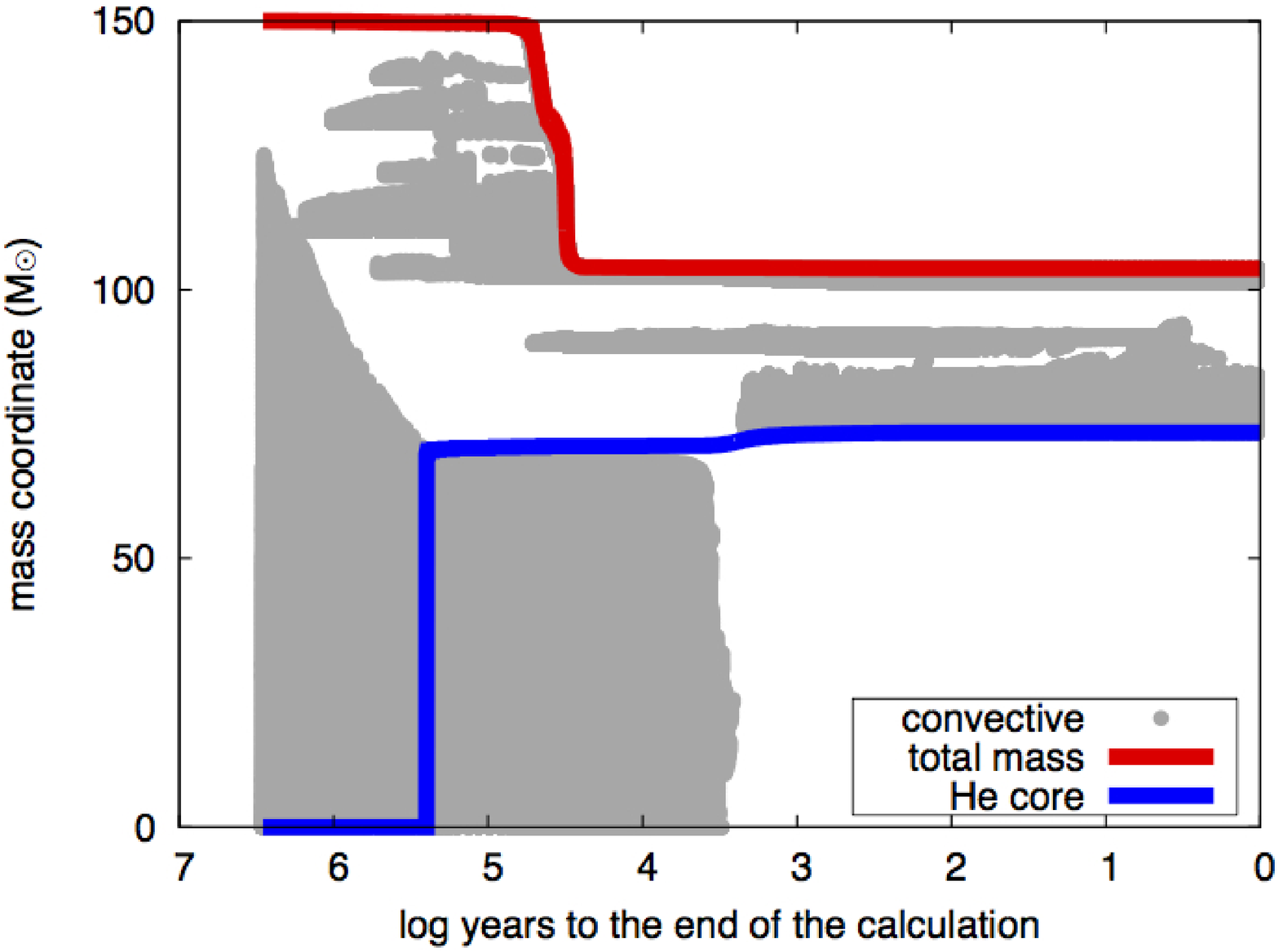}\\
\includegraphics[width=\columnwidth]{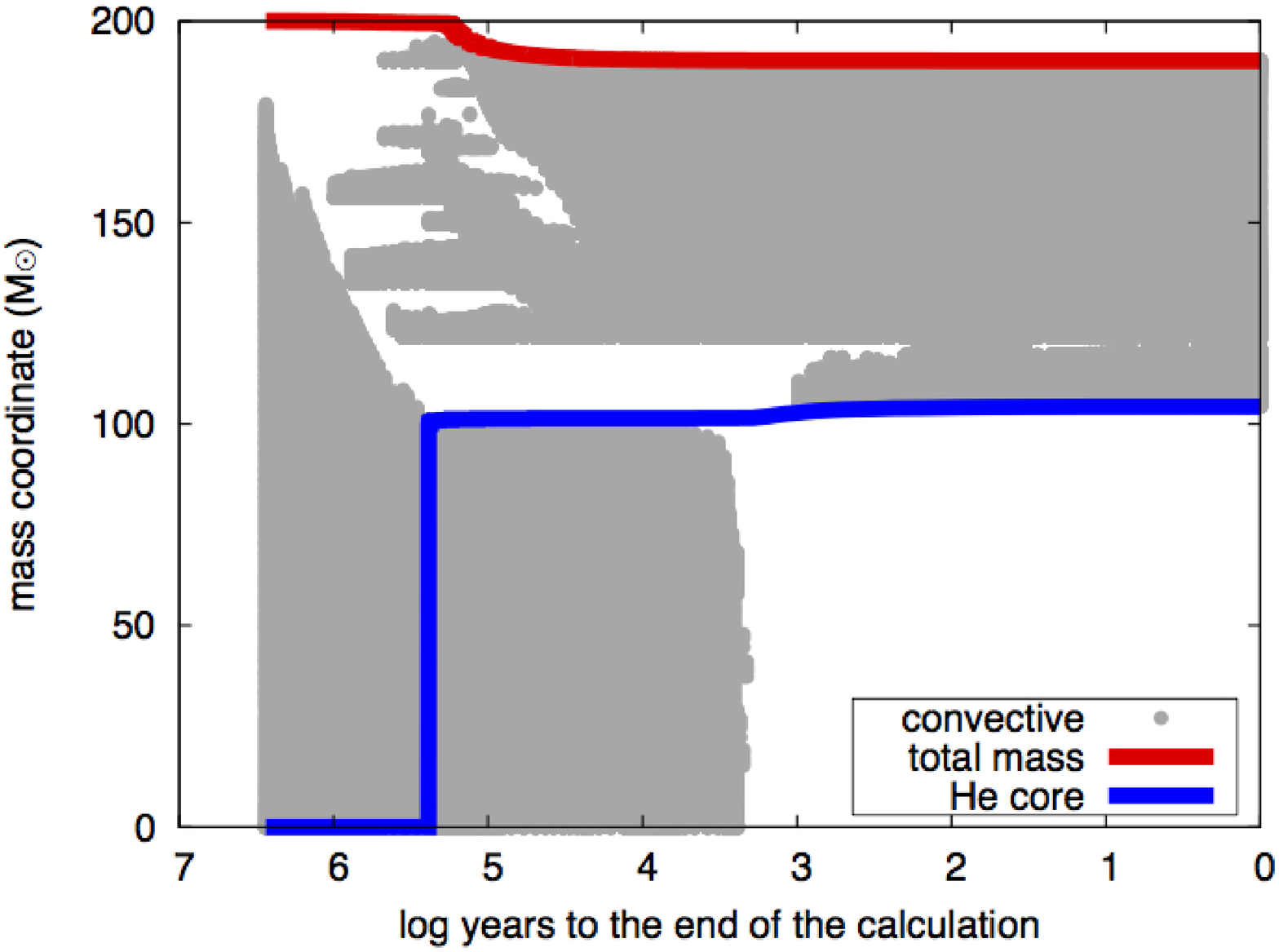}
\includegraphics[width=\columnwidth]{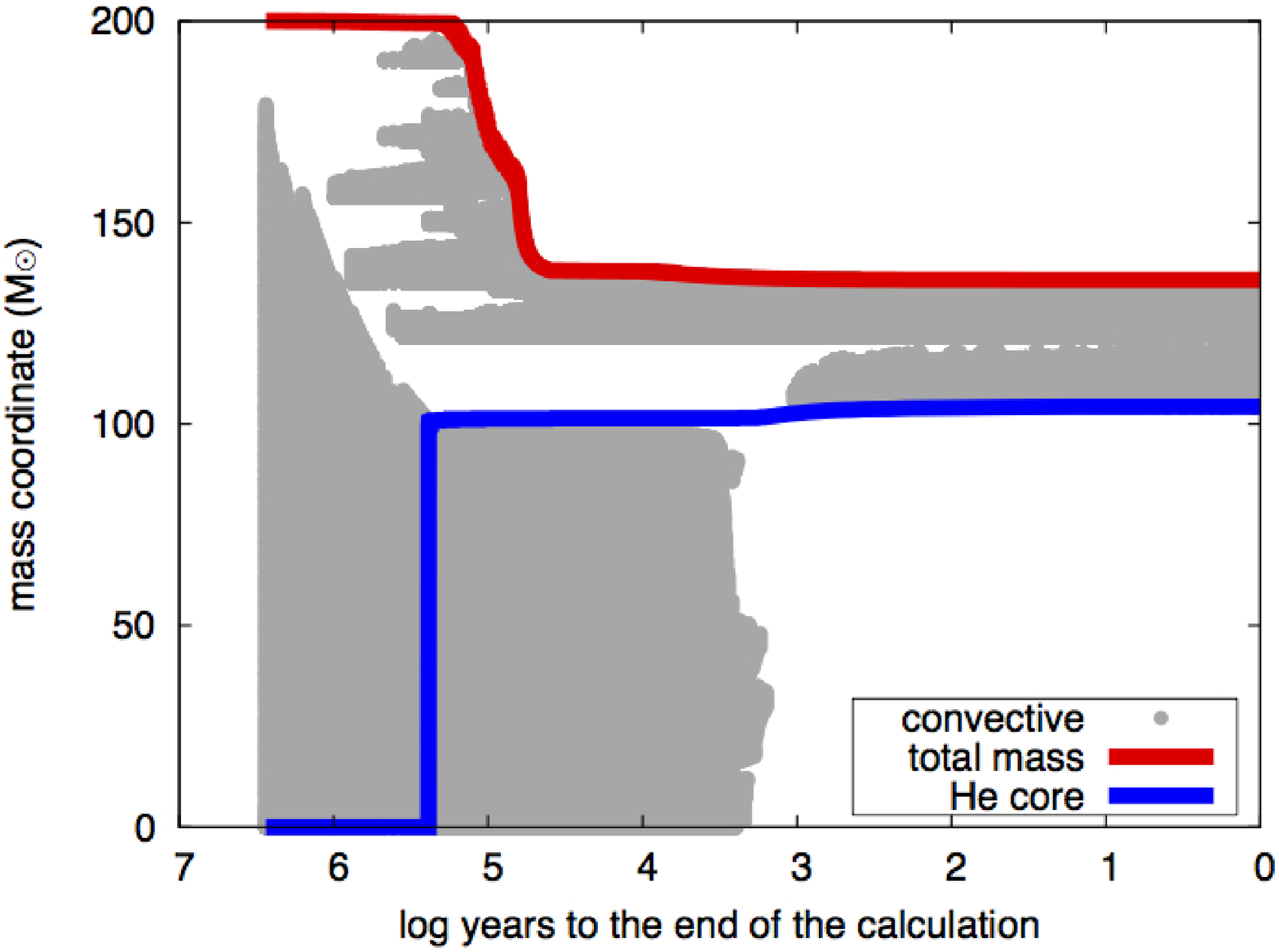}\\
\includegraphics[width=\columnwidth]{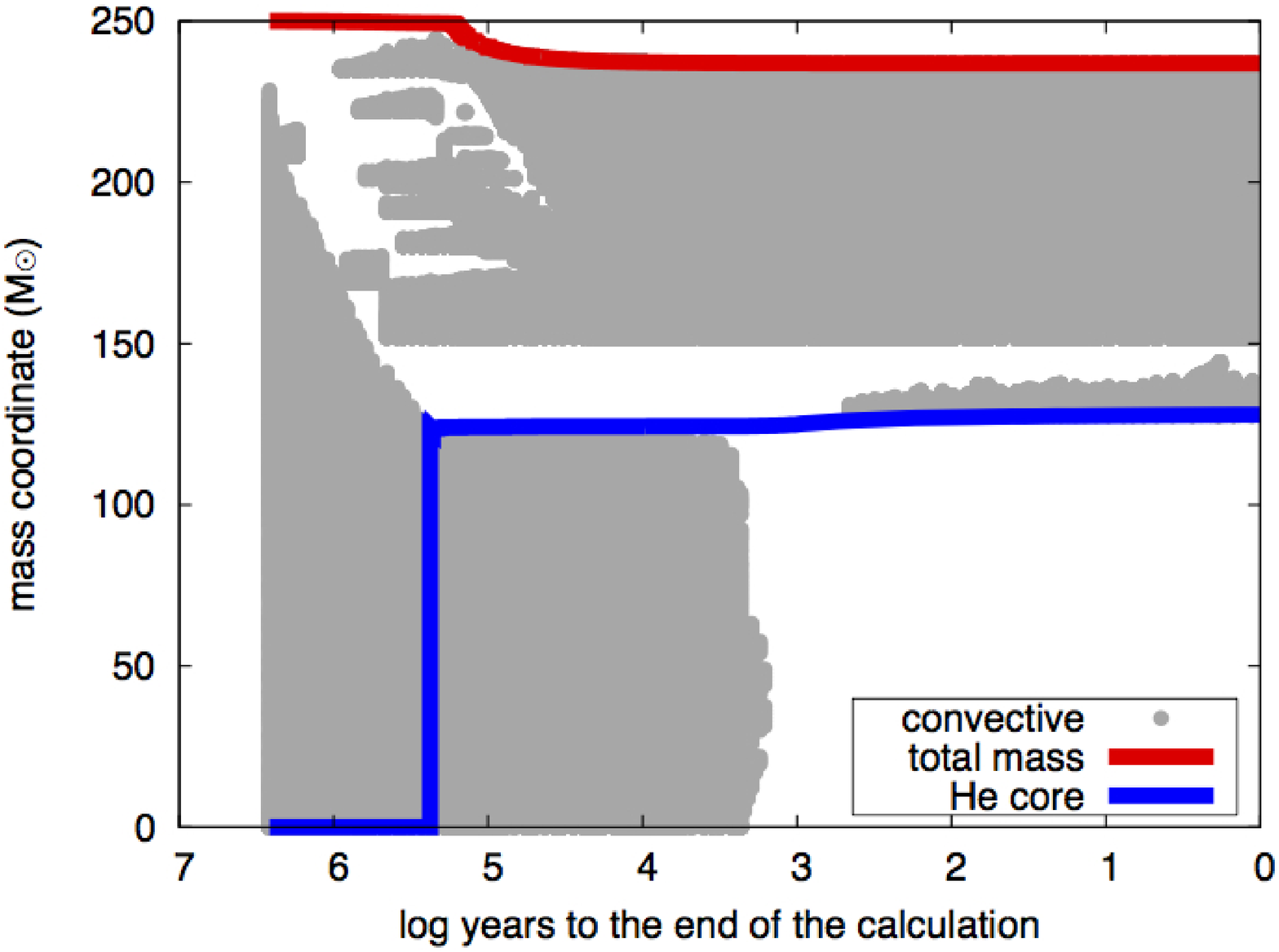}
\includegraphics[width=\columnwidth]{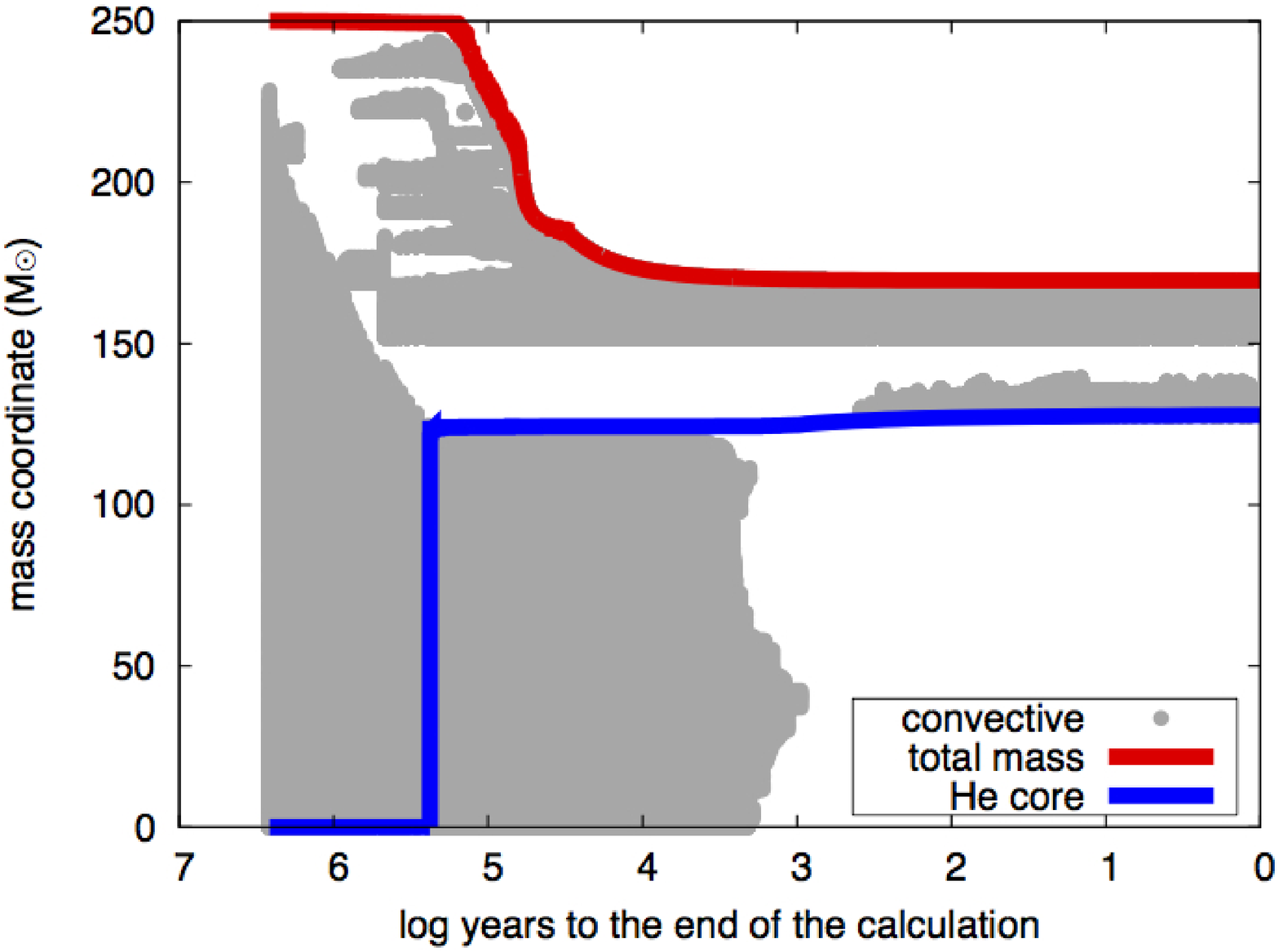}
\caption{
Kippenhahn diagrams for our stellar evolution models.
Models in the left column are computed without pulsation-driven mass loss ($\varepsilon=0$)
and those in the right column include pulsation-driven mass loss
 with $\varepsilon=0.1$.
The initial masses of the models are 150 (top), 200 (middle),
and 250~\Msun\ (bottom).
}
\label{kippen}
\end{figure*}

\begin{figure*}
\centering
\includegraphics[width=\columnwidth]{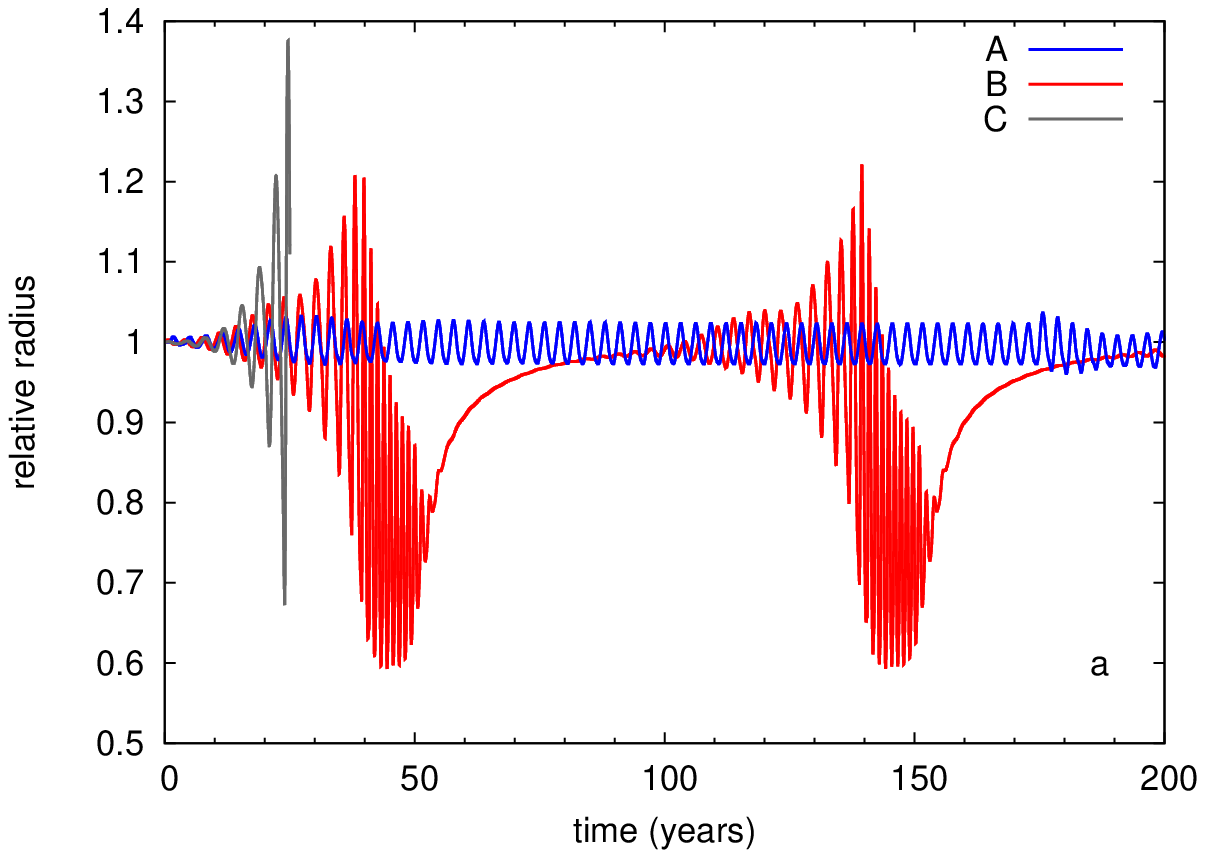}
\includegraphics[width=\columnwidth]{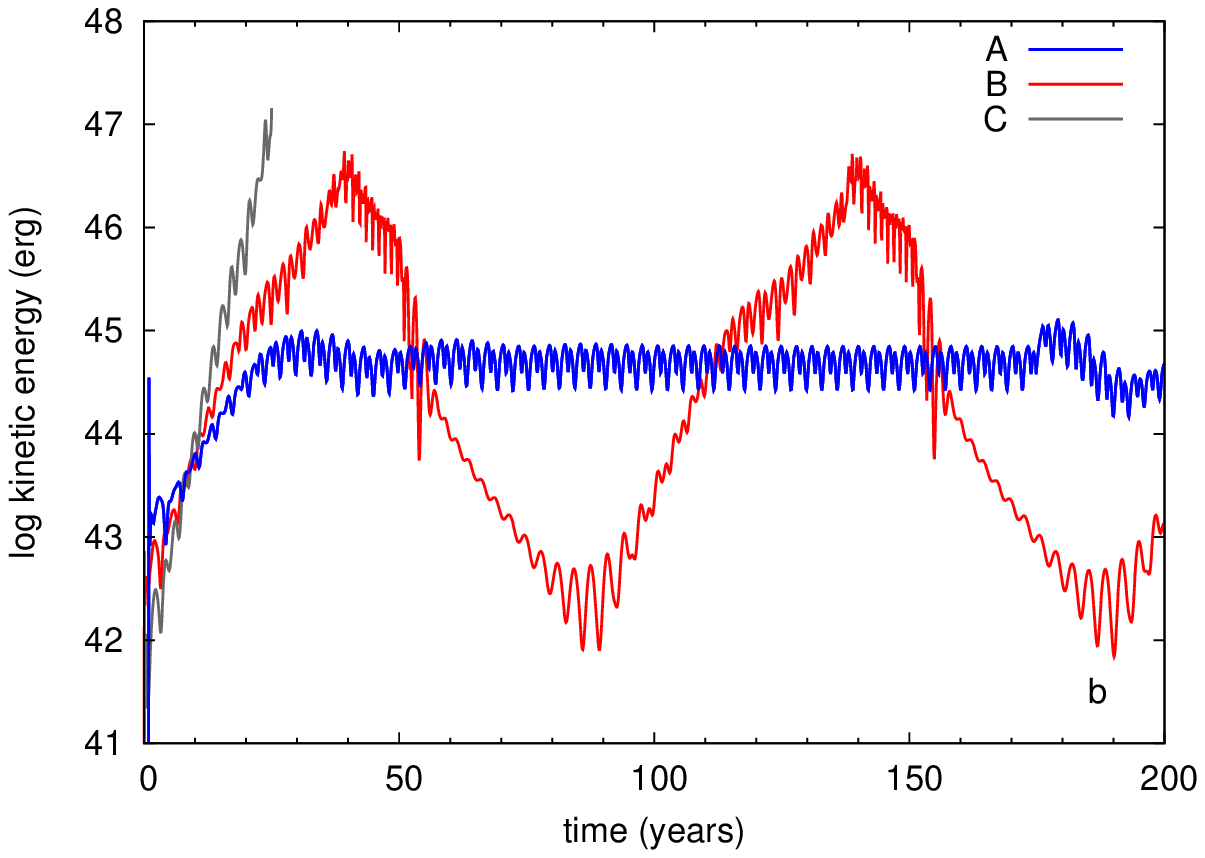} \\
\includegraphics[width=\columnwidth]{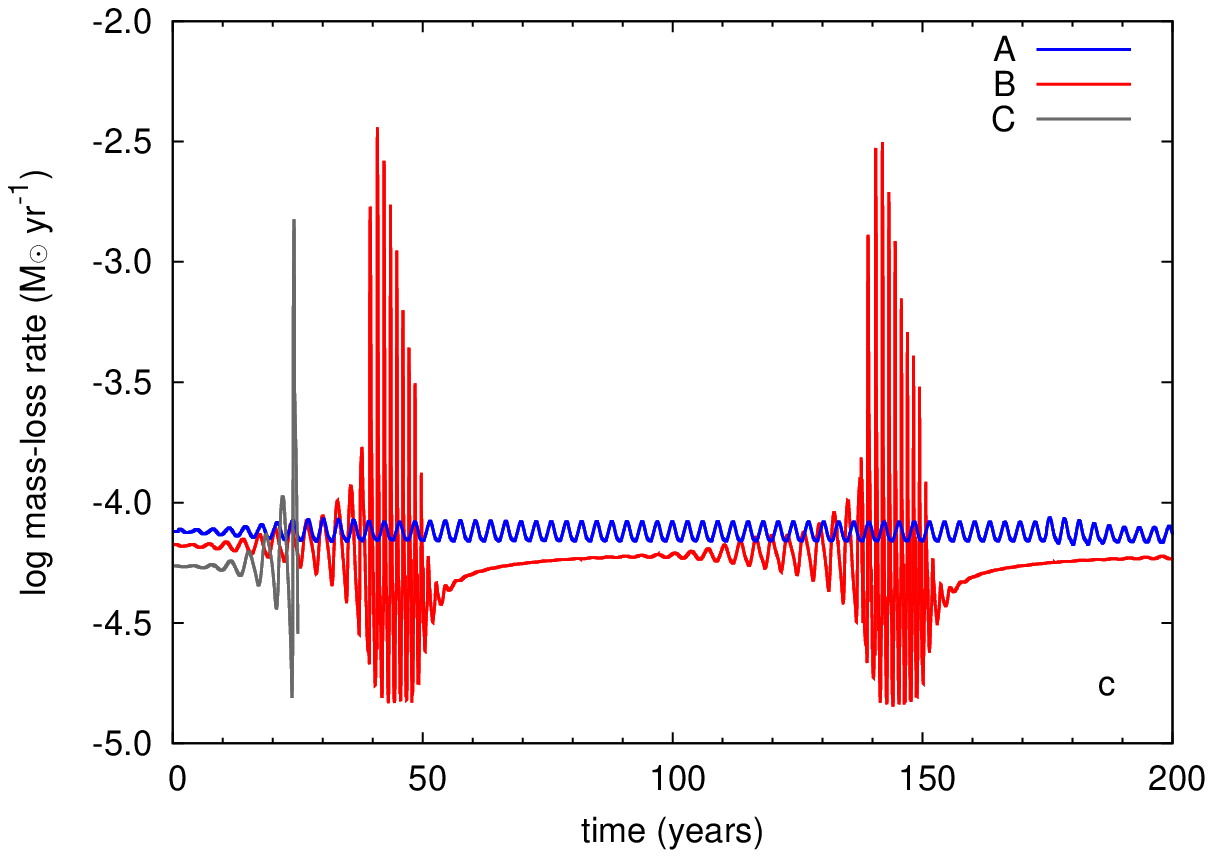}
\includegraphics[width=\columnwidth]{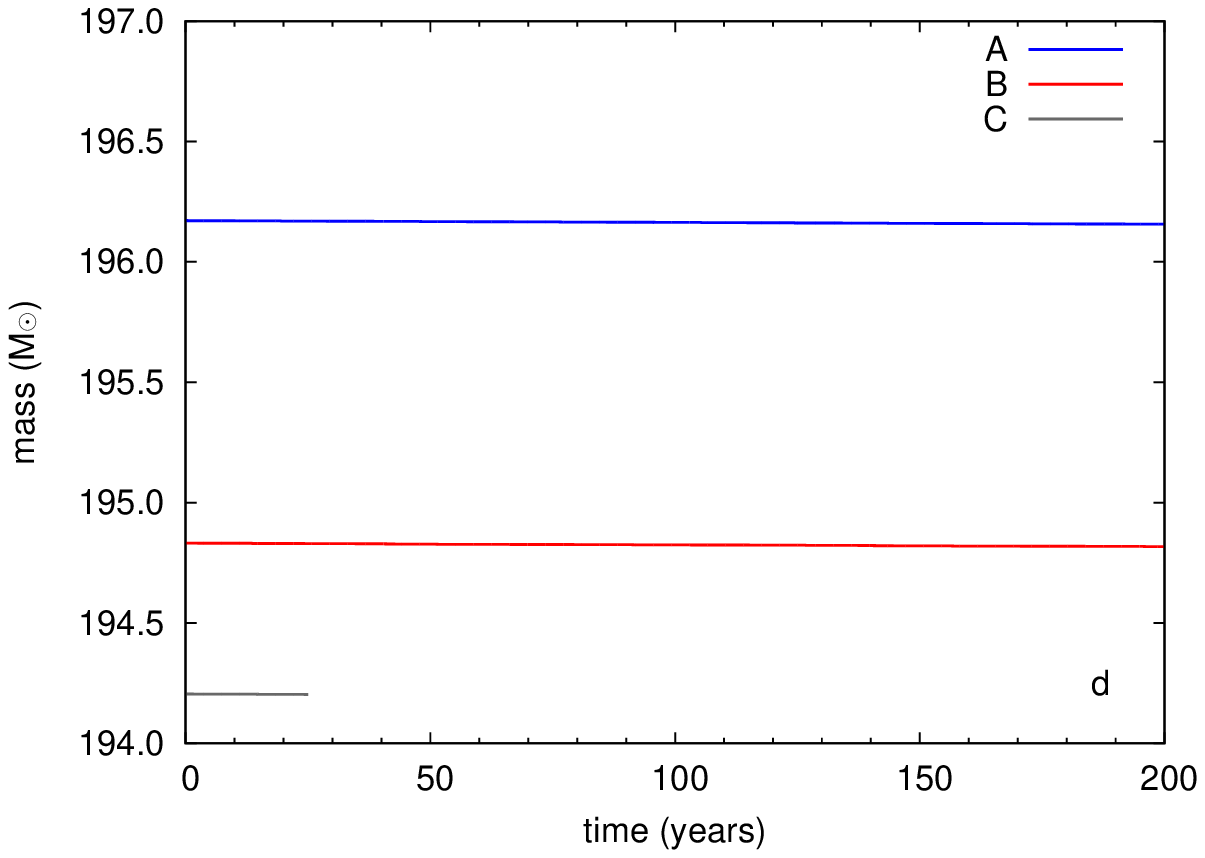}
\caption{
Examples of the pulsations of the PISN progenitors during the RSG stage.
Three representative 200~\Msun\ models are shown.
No pulsation-induced mass-loss enhancement is taken into account in the models of this figure.
Evolution of (a) relative surface radii, (b) total kinetic energy in the
 star, (c) mass-loss rates, and (d) total stellar masses is shown.
}
\label{standard}
\end{figure*}

\section{Stellar evolution}\label{evolution}
We follow evolution of stellar structure by using a public stellar evolution code
\texttt{MESAstar} version 6208 \citep{paxton2011,paxton2013}.
We use the hydrodynamic mode of the code throughout this paper.
We follow the evolution of three PISN progenitors whose initial masses
are 150, 200, and 250 \Msun. The initial metallicity is $Z=0$.
\texttt{MESAstar} is demonstrated to be able to follow the RSG pulsations
that we investigate in this paper \citep{paxton2013}.
An example of input parameters for \texttt{MESAstar} used in
this study is shown in Appendix \ref{appendix}.

The mixing-length theory is adopted to treat convection using
the Schwarzschild criterion.
The mixing-length parameter is set as 1.6.
Overshooting and semi-convection are not taken into account.
We use the `approx21' nuclear network which covers the major important nuclear
reactions \citep{timmes1999}.
Although our stars are metal-free at the beginning,
the stellar surface metallicity slightly increases as the stars evolve.
As a result, the stellar mass is reduced due to the radiation-driven wind, especially
during the post-main-sequence.
When the stellar effective temperature is higher than $10^4$~K,
we use the mass-loss rates formulated by \citet{vink2001}.
For the cooler stars, we use the mass-loss rates of
\citet{dejager1988} with the metallicity dependence of $Z^{0.5}$ \citep{kudritzki1987}.

The stars are evolved from the pre-main-sequence stage.
The calculations are terminated soon after the stellar center starts to
contract as a consequence of the pair instability (Fig.~\ref{central}).
We first follow the stellar evolution 
without taking the effect of the RSG pulsations into account.
The stellar properties at the end of these calculations are listed in
Table~\ref{tablestandard} (the $\varepsilon=0$ models).
The evolution of our stars in the
Hertzsprung-Russell (HR) diagram is presented in Fig.~\ref{hrd}.
All the stars evolve into RSGs and explode during the RSG stage as PISNe.
The evolutionary tracks of our PISN progenitors are consistent with
those obtained in previous studies \citep[e.g.,][]{yoon2012}.
The Kippenhahn diagrams for our models are shown in Fig.~\ref{kippen}.
Our models without mass-loss enhancement have large convective
hydrogen-rich envelopes during the late stages when they are RSGs.
We investigate the pulsational properties of the stars at the RSG stage
in the next section.

\section{Pulsations and mass loss}\label{pulsation}
\subsection{Pulsation properties}\label{standardpulsation}
\begin{figure}
\centering
\includegraphics[width=\columnwidth]{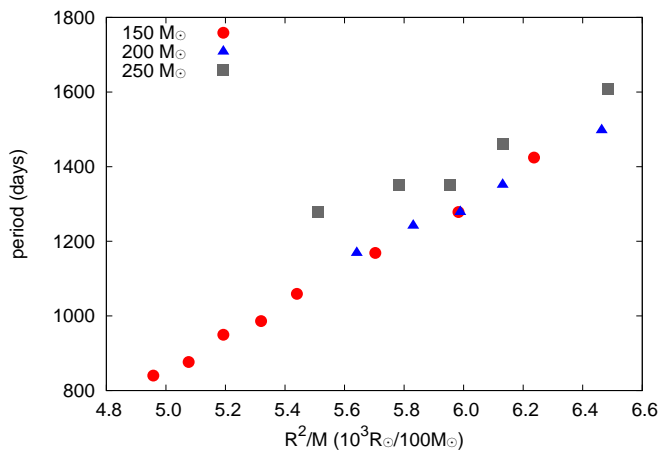}
\caption{
Period $P$ of the pulsations we obtained in our RSGs as a function of
$R^2/M$.
}
\label{period}
\end{figure}

To investigate the pulsational properties of the PISN progenitors
during the RSG stage,
we use the stellar models that are indicated in Fig.~\ref{hrd}.
We restart the calculations from these models 
by forcing the time steps to be less than 0.001 years to follow the
pulsations of the stars.
The pulsational periods of RSGs were found to be of the order of 1000
days in the previous studies (\citealt{heger1997,yoon2010}, see also Fig.~\ref{period})
and we choose small enough time steps to resolve such presumed periods.

We follow the evolution with the small time steps for at least 100 years
to check whether pulsations develop and if so, whether the pulsational amplitude grows.
The models in which we do not find pulsations with a growing amplitude
are indicated with open circles in Fig.~\ref{hrd}.
These models are either pulsationally stable, or they are 
unstable but with such a small growth rates that the pulations are
damped numerically. In either case, we assume that pulsationally
induced mass loss may be neglected in these models.

For the models marked with filled symbols in Fig.~\ref{hrd},
we find pulsations with growing amplitudes, as in previous studies
of less massive RSGs \citep[e.g.,][]{heger1997,yoon2010}.
We assume the convective flux to adjust instantly during the pulsations,
as \citet{heger1997,yoon2010},
even though the convective and the pulsation timescales are comparable.
However,
note that \citet{heger1997} found that the pulsational growth rates
in their non-linear numerical calculations
to be consistent with those computed from linear stability analysis,
where the convective flux is assumed to be frozen in.
Since \citet{heger1997} found that two different extreme assumptions on
the behavior of the convective flux during the pulsation lead to very
similar results, and also \citet{langer1971} found that the growth rate
of the pulsations does not depend on the phase-lag parameter in
the simple time-dependent convection model of \citet{arnett1969},
we are confident in the main features of our pulsational analysis.
Still, a physically more correct treatment of the coupling between
convection and pulsations is desirable, which proves to be difficult
to be set up in a parameter-free way \citep{gastine2011,sonoi2014}.

We find three different patterns in the way how the amplitudes grow (Fig.~\ref{standard}).
In the first pattern named A, the pulsational amplitude grows
exponentially at first. Then, the pulsation saturates
and the stars continue to pulsate with a constant amplitude.
The pulsational amplitudes of the stars with the second pattern (B) grow
exponentially at first but the amplitudes are suddenly damped.
Then, the amplitudes start to grow exponentially again until they are
damped again. The exponential growth and the sudden damping of
the pulsational amplitudes are repeated.
The likely reason for the sudden damping is the increase of
the radiation-driven mass loss (Fig.~\ref{standard}c).
When the pulsational amplitude gets sufficiently
large, the radiation-driven mass-loss rates become large because the stars get cooler as
they expand. 
When the amplitude is damped, the mass-loss rates become
small again and pulsations can grow again.
In the final pattern C,
the pulsational amplitudes grow exponentially until the stellar surface reaches the escape
velocity. The time steps of our calculations become very small when 
this happens and we stop the calculations at that moment.
The pulsational patterns we obtained for each models are indicated with
different symbols in Fig.~\ref{hrd}.

\begin{figure}
\centering
\includegraphics[width=\columnwidth]{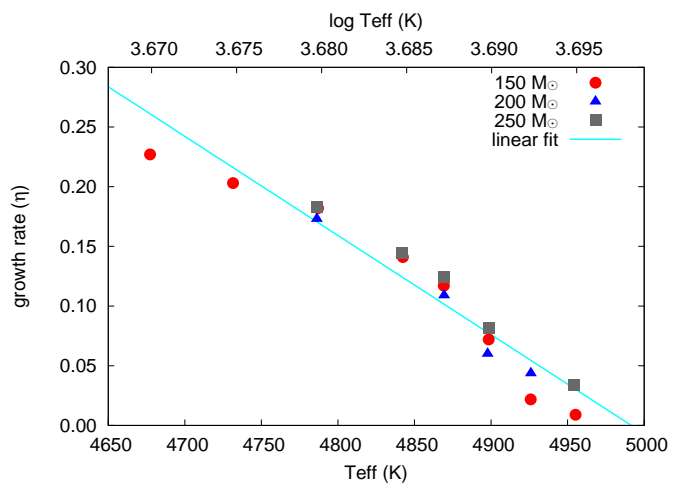}
\caption{
Growth rate $\eta$ as a function of \Teff.
The growth rate linearly correlates with \Teff\ (Eq.~\ref{eqeta}).
}
\label{Teff-eta}
\end{figure}

In the unstable models, we can find
the exponential growth of pulsational amplitudes at least initially
in all the patterns.
We evaluate the period $P$ and the growth rate $\eta$ during the initial
exponential growth phase for these unstable models.
If the exponential growth of the pulsations
is proportional to $\exp(i\sigma t)$,
the growth rate is defined as $\eta\equiv-\Im\sigma/\Re\sigma$
where $\Im\sigma$ is the imaginary part of $\sigma$ and
$\Re\sigma$ is the real part of $\sigma$.
$\Re\sigma$ and $\Im\sigma$ can be estimated from $P$ and the
exponential growth of the surface radius, respectively.
The periods in our pulsating models are summarized in Fig.~\ref{period}.
The periods of the pulsations in convective stars are expected to follow
$P\propto R^2/M$ \citep{gough1965} as we find in our models,
where $R$ is the stellar radius.
The growth rates $\eta$ are presented in Fig.~\ref{Teff-eta}.
We find the following strong correlation between $\eta$ and the effective
temperature \Teff:
\begin{equation}
 \eta = (-8.30\pm0.59)\times 10^{-4} T_\mathrm{eff} + (4.15\pm0.29),\label{eqeta}
\end{equation}
where \Teff\ is in K and the error is the standard error.
The obtained relation indicates that the stars with $\Teff\lesssim 4992$ K
are pulsationally unstable in our models.

\begin{figure}
\centering
\includegraphics[width=\columnwidth]{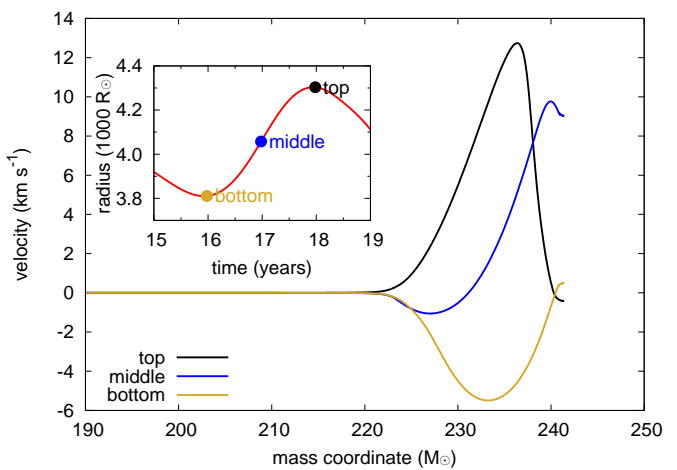}
\caption{
Internal velocity structure of a pulsating model.
The velocity structure at three representative times
indicated in the inset are shown.
The model is obtained from the 250~\Msun\ sequence at
 $\log\Teff/\mathrm{K}=3.680$ and $\log L/L_\odot=6.872$ with $\varepsilon=0$.
The model is the same as that with $\varepsilon=0$ presented in Fig.~\ref{enhanced}.
}
\label{velocity}
\end{figure}

\begin{figure*}
\centering
\includegraphics[width=\columnwidth]{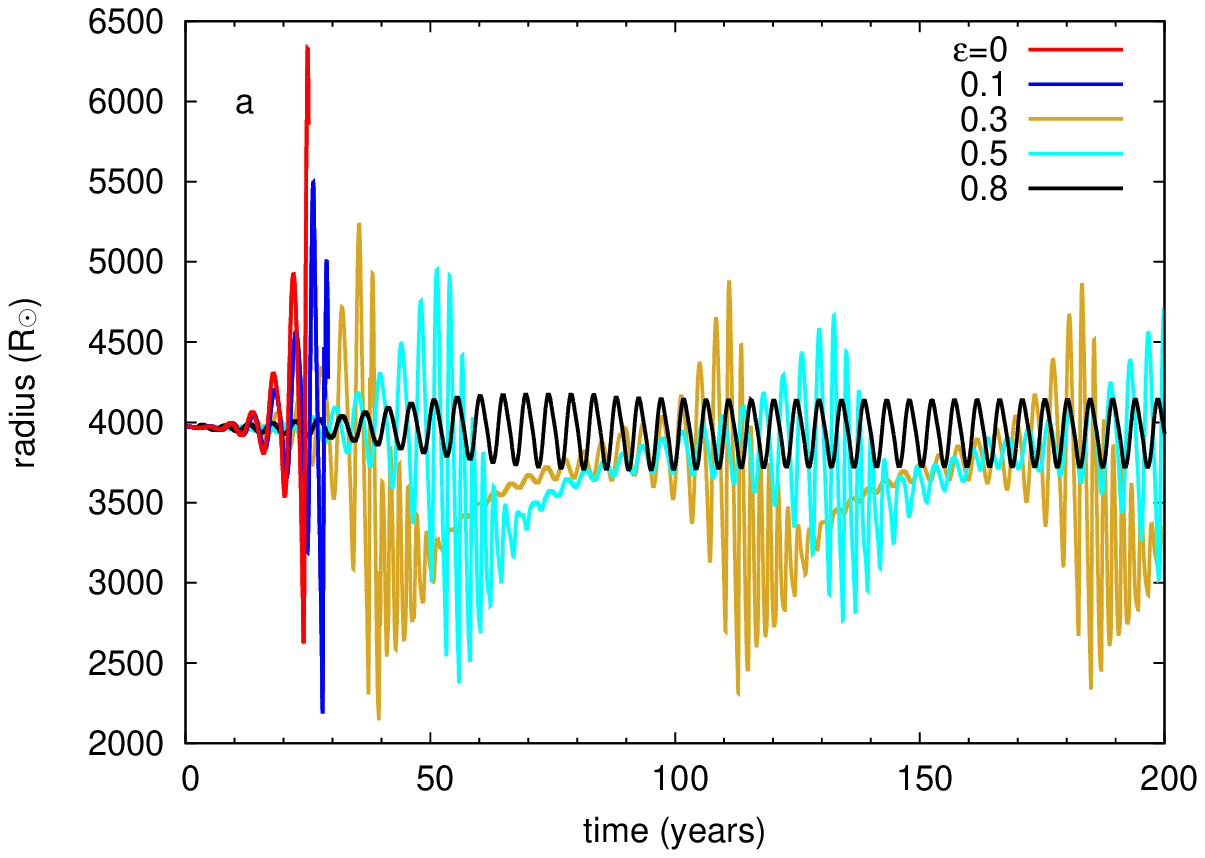}
\includegraphics[width=\columnwidth]{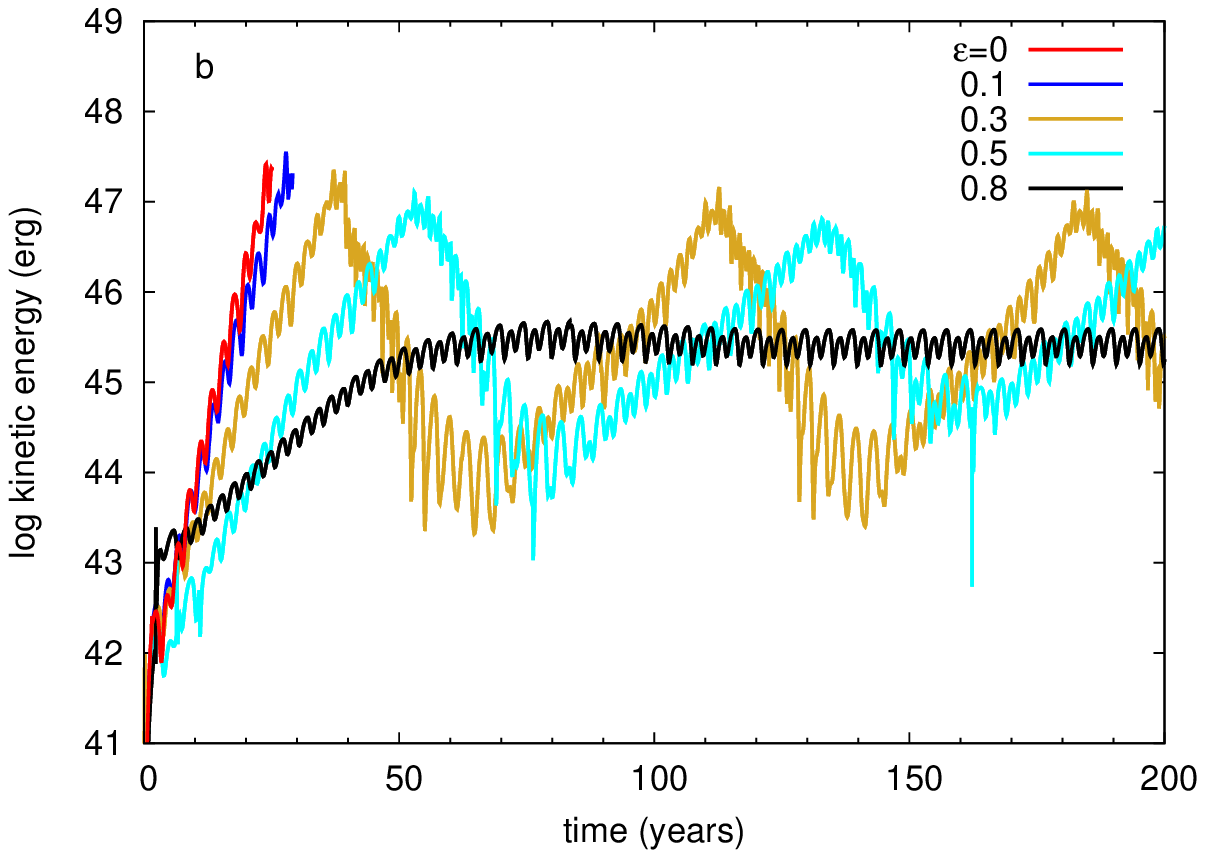} \\
\includegraphics[width=\columnwidth]{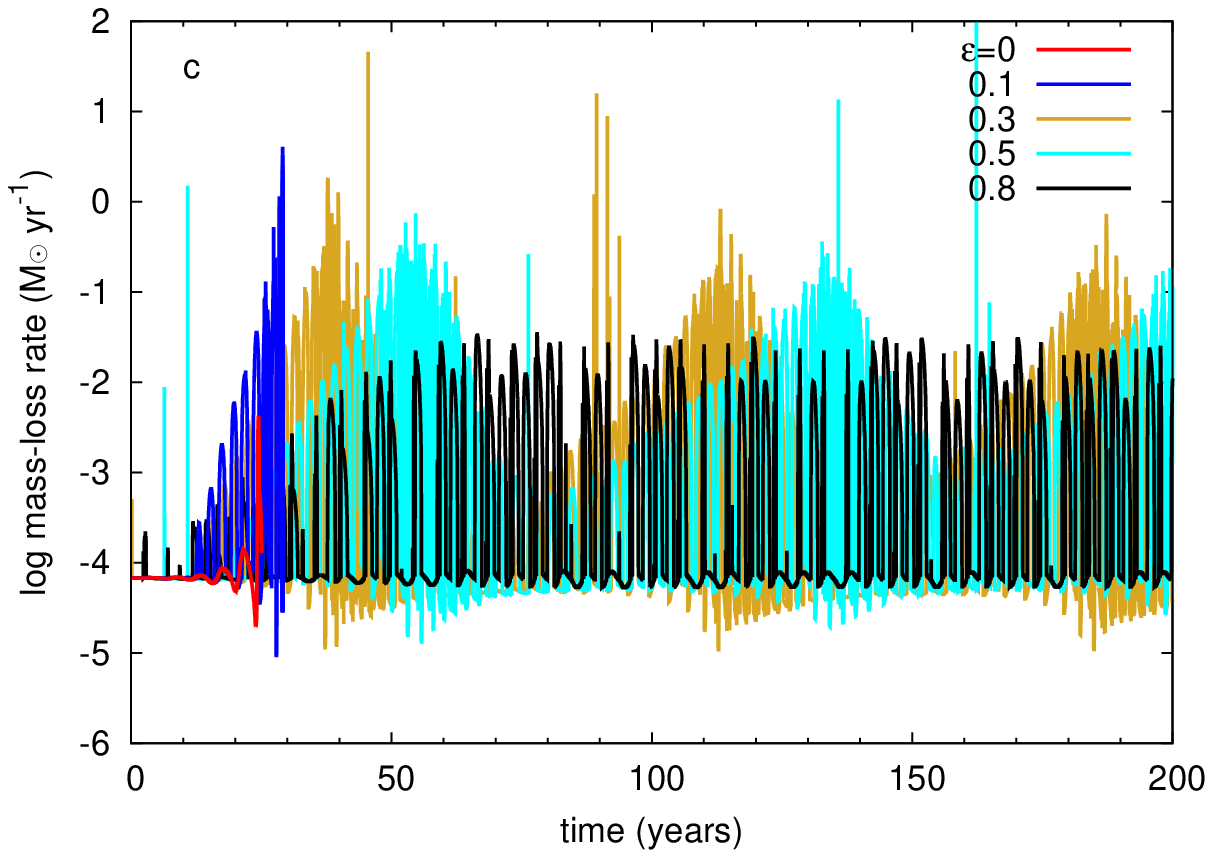}
\includegraphics[width=\columnwidth]{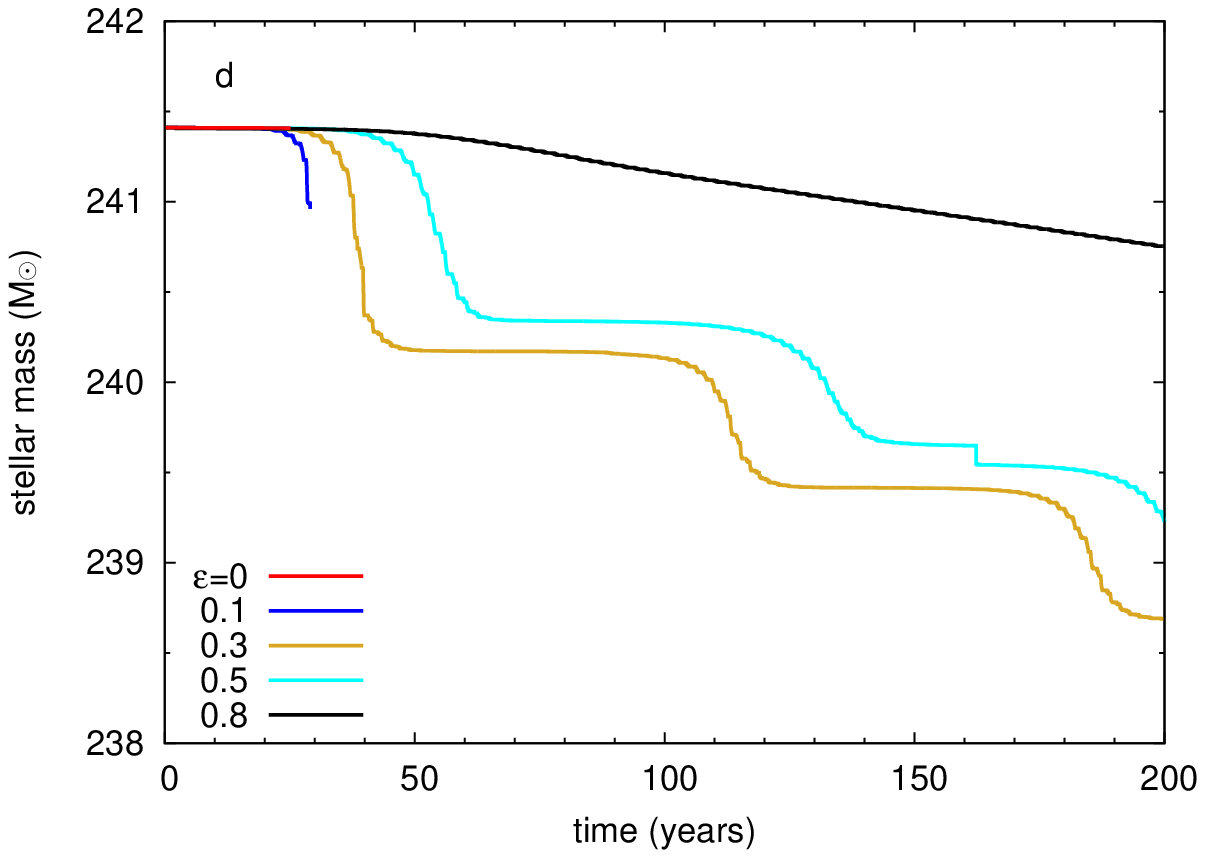}
\caption{
Examples of the evolution of the RSG pulsations
in which the mass-loss enhancement by
the pulsations is taken into account. A fraction $\varepsilon$ of 
the gained kinetic energy is assumed to be used to induce mass loss.
All the models are started from the 250~\Msun\ model at $\log\Teff/\mathrm{K}=3.680$ and $\log L/L_\odot=6.872$.
The model with $\varepsilon=0$ is calculated without mass-loss enhancement.
Evolution of (a) surface radii, (b) total kinetic energy in the star,
(c) mass-loss rates, and (d) total stellar masses is shown.
}
\label{enhanced}
\end{figure*}

\subsection{Pulsation-induced mass loss}\label{sec:enhanced}
We have investigated the pulsational properties of the PISN progenitors
during their RSG stage. So far, we have not related the pulsations
to stellar mass loss.
As we introduced in Section \ref{introduction},
stellar pulsations may be able to induce mass loss.
In this section, we relate the pulsational properties of RSGs to the
stellar mass-loss rates. The consequences of 
pulsation-driven mass loss are investigated in the next section.

We relate the kinetic energy gain of the growing pulsations
to mass loss. We assume that a fraction $\varepsilon$ of the gained
kinetic energy $\Delta E_\kin$ at each time step is used to induce mass loss.
We estimate the mass-loss rate $\dot{M}_\kin$ induced by the
pulsations as
\begin{equation}
 \dot{M}_\kin = \frac{2\varepsilon\Delta E_\kin}{v_\esc^2 \Delta t},
\label{Mdotkin}
\end{equation}
where $v_\esc$ is the escape velocity from the stellar surface
and $\Delta t$ is the time step \citep[cf.][]{appenzeller1970b,baraffe2001}.
In the following stellar evolution calculations,
we account of the pulsation-induced mass loss (Eq.~\ref{Mdotkin}) in addition
to the radiation-driven mass loss.
The kinetic energy does not always increase even if the pulsational amplitude is
exponentially growing (cf. Fig.~\ref{standard}b).
When $\Delta E_\kin<0$, $\dot{M}_\kin$ is set to 0.

When $\dot{M}_\kin>0$, a fraction of the kinetic energy in the star is supposed
to initiate mass loss and the corresponding amount of the
kinetic energy needs to be reduced from the pulsations.
When the star with the kinetic energy $E_\kin$ gains $\Delta E_\kin$, then
$\varepsilon \Delta E_\kin$ is supposed to initiate mass loss
and the same amount of the kinetic energy in the star needs to be reduced.
We reduce the kinetic energy in the stars by reducing the velocities of
all the mesh points in the stars as
\begin{equation}
 v_\mathrm{new}=v_\mathrm{old}\sqrt{\frac{E_\kin+\Delta E_\kin
  -\varepsilon \Delta E_\kin}{E_\kin+\Delta E_\kin}}, \label{eqvel}
\end{equation}
where $v_\mathrm{new}$ and $v_\mathrm{old}$ are the new and old
velocities, respectively.
By adopting $v_\mathrm{new}$, the kinetic energy which is supposed to induce mass
loss is actually removed from the stellar models.

We obtain $E_\kin$, and thus $\Delta E_\kin$, by integrating the kinetic
energy of all the mass shells in the stars. Then, we reduce the
velocities inside the entire star by Eq.~(\ref{eqvel}).
However, the pulsating layers gaining the kinetic energy 
are only located near the surface, as can be seen in the velocity
structure of an unstable star presented in Fig.~\ref{velocity}.
Although we obtain and reduce the kinetic energy in the entire star,
only the outer layers which induce the mass loss have an essential effect on
the determination of $E_\kin$ and the reduction of the velocities.

\subsection{Pulsation-induced mass-loss rates}
Using the mass-loss prescription with the kinetic energy reduction
explained in the previous section,
we perform evolutionary calculations with small time
steps ($\Delta t \leq 0.001$ years) as in Section
\ref{standardpulsation} to investigate the effect of the
pulsation-driven mass loss on the evolution of the stars.
For this purpose, we select some pulsationally unstable models
obtained in Section \ref{standardpulsation} and calculate the
stellar evolution with $\varepsilon=0.1,0.3,0.5,$ and $0.8$.

Fig.~\ref{enhanced} shows examples of the results with the mass-loss enhancement.
All the models presented in Fig.~\ref{enhanced} are evolved starting from
the 250~\Msun\ model with $\log\Teff/\mathrm{K}=3.680$ and $\log L/L_\odot=6.872$.
The model with $\varepsilon=0$ is the original model without the extra mass loss.
The original model has the pulsation pattern C.
The models with mass-loss enhancement are indicated with
the adopted $\varepsilon$.
Since a fraction of the gained kinetic energy is taken out for the mass loss
in the new models,
their growth rates are smaller than those without the mass-loss
enhancement. This can be clearly seen in Fig.~\ref{enhanced}b
where the total stellar kinetic energy of the models is shown.
The changes in the growth rates result in changes of the pulsation
patterns. While the original model ($\varepsilon=0$)
and the model with $\varepsilon=0.1$ have the pattern C,
the enhanced models with $\varepsilon=0.3$ and 0.5 show the pattern B.
In the $\varepsilon=0.8$ model, the exponential growth of the
pulsational amplitude is terminated early on and the pulsational pattern
becomes A. This is because most of the gained kinetic energy is used for
the mass loss and the amplitude cannot grow much.

The mass-loss rates of our models are presented in Fig.~\ref{enhanced}c.
The pulsation-driven mass loss is activated only when kinetic
energy is gained by the pulsations,
making the mass-loss rates strongly time dependent.
The resulting mass evolution is shown in Fig.~\ref{enhanced}d.
As the pulsation grows, the mass-loss rates increase due to the increase
of the energy gain of the pulsations.
A significant amount of mass is lost when the mass-loss rates become large enough.
In the cases with $\varepsilon=0.3$ and 0.5, the large mass ejection results
in damping of the pulsation. The pulsation amplitude grows again after the
damping and large mass ejections occur intermittently.
In the case of $\varepsilon=0.8$, the amplitude growth stops at early time
because a large amount of the gained kinetic energy is used for the mass loss,
making the growth less efficient. In this case, the kinetic energy
induced mass loss also saturates.

Based on the pulsational calculations with the enhanced mass loss,
we estimate the average mass-loss rates of each models.
If the pulsational pattern is A, we obtain the average mass-loss rate
when the pulsation is saturated.
For instance, for the model with $\varepsilon=0.8$ in
Fig.~\ref{enhanced}, the estimated mass-loss rate is $3\times 10^{-3}$ \mrunit.
For the pattern B, the representative mass-loss rates are obtained
based on the mass lost after the first large mass ejection.
For example, in the model with $\varepsilon=0.3$ in Fig.~\ref{enhanced},
we take the mass just after the large mass ejection at 50 years
and we obtain the average mass-loss rate in 50 years, namely,
$2\times10^{-2} M_\odot~\mathrm{yr^{-1}}$. For the pattern C, we
evaluate the mass difference between the initial and final models
and divide it by the time that we could follow the model to
obtain the average mass-loss rates.
These ways to estimate the average mass-loss rates are not free of
arbitrariness and
the estimated mass-loss rates can change by a small factor depending on the adopted ways.

\begin{figure}
\centering
\includegraphics[width=\columnwidth]{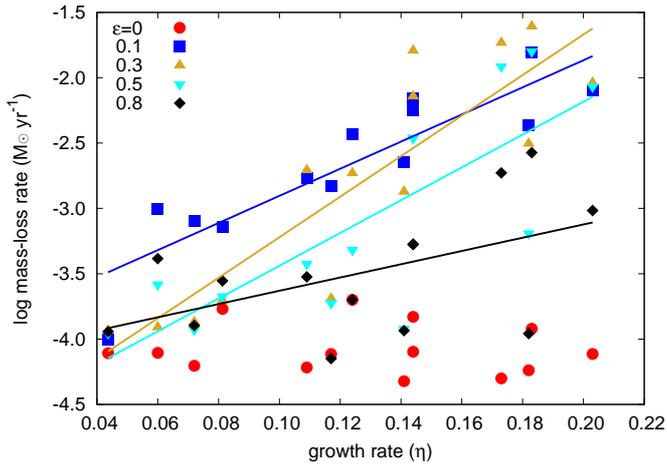}
\caption{
Average mass-loss rates of the stars with the pulsation-induced mass
 loss. The mass-loss rates depend on the growth
 rate $\eta$. Exponential fits to the estimated mass-loss rates (Eq. \ref{eqenhancedmasslossrate})
are shown with lines.
}
\label{eta-massloss}
\end{figure}

The average mass-loss rates we obtained are shown
as a function of the growth rate in Fig.~\ref{eta-massloss}.
The growth rates are those obtained in the models without the mass-loss
enhancement $(\varepsilon=0)$.
We use the growth rate without the mass-loss enhancement so that we can
relate the enhanced mass-loss rates to \Teff\ with
Eq.~(\ref{eqeta}), as is applied in the next section.
We find that the mass-loss rates 
with the pulsations generally increase as the growth rates increase.
We find substantial mass-loss rates in the models with
$\varepsilon=0.1,0.3,$ and 0.5, especially when $\eta$ is large.
For the models with $\varepsilon=0.8$, the 
mass-loss rates are not as large as those in other enhanced
models. This is presumably because of
the large reduction of the pulsational amplitudes due to the large loss of the
kinetic energy.
Note that for $\varepsilon=1$, the pulsational mass loss must vanish.

By fitting the mass-loss rates in Fig.~\ref{eta-massloss} with
exponential functions,
we find the following pulsation-induced mass-loss rates:
\begin{eqnarray}
\log_{10} \dot{M}_\mathrm{kin}=\left\{ \begin{array}{ll}
(10.38 \pm 1.58)\eta -(3.94\pm 0.21) & (\varepsilon=0.1),\\ 
(15.51 \pm 2.92)\eta -(4.77\pm 0.33) & (\varepsilon=0.3),\\ 
(12.55 \pm 2.88)\eta -(4.69\pm 0.39) & (\varepsilon=0.5),\\ 
( 5.08 \pm 2.87)\eta -(4.14\pm 0.33) & (\varepsilon=0.8),\\ 
\end{array} \right. \label{eqenhancedmasslossrate}
\end{eqnarray}
where $\dot{M}_\mathrm{kin}$ is in the unit of $M_\odot~\mathrm{yr^{-1}}$.


\section{Consequences of the mass-loss enhancement}\label{consequences}
\subsection{Stellar evolution}
In the previous section, we have selfconsistently evaluated the effect of
pulsationally induced mass loss on the pulsations and on the evolution of the stars.
However, these calculations are time consuming and hard to be conducted over more
than a few hundred pulsation cycles. 
To investigate the effect of the enhanced mass loss 
on the entire evolution and final fates of PISN progenitors, we calculate the evolution of the
150, 200, and 250~\Msun\ stars with large time steps such that 
pulsations cannot develop, and
we add the pulsation-induced mass-loss rate $\dot{M}_\mathrm{kin}$
by applying Eq.~(\ref{eqenhancedmasslossrate}).
The growth rate $\eta$ is evaluated based on $T_\mathrm{eff}$
through Eq.~(\ref{eqeta}). When $\eta\leq0$, we set $\dot{M}_\mathrm{kin}=0$.
We investigate the cases of $\varepsilon=0.1$ and 0.3.

\begin{figure}
\centering
\includegraphics[width=\columnwidth]{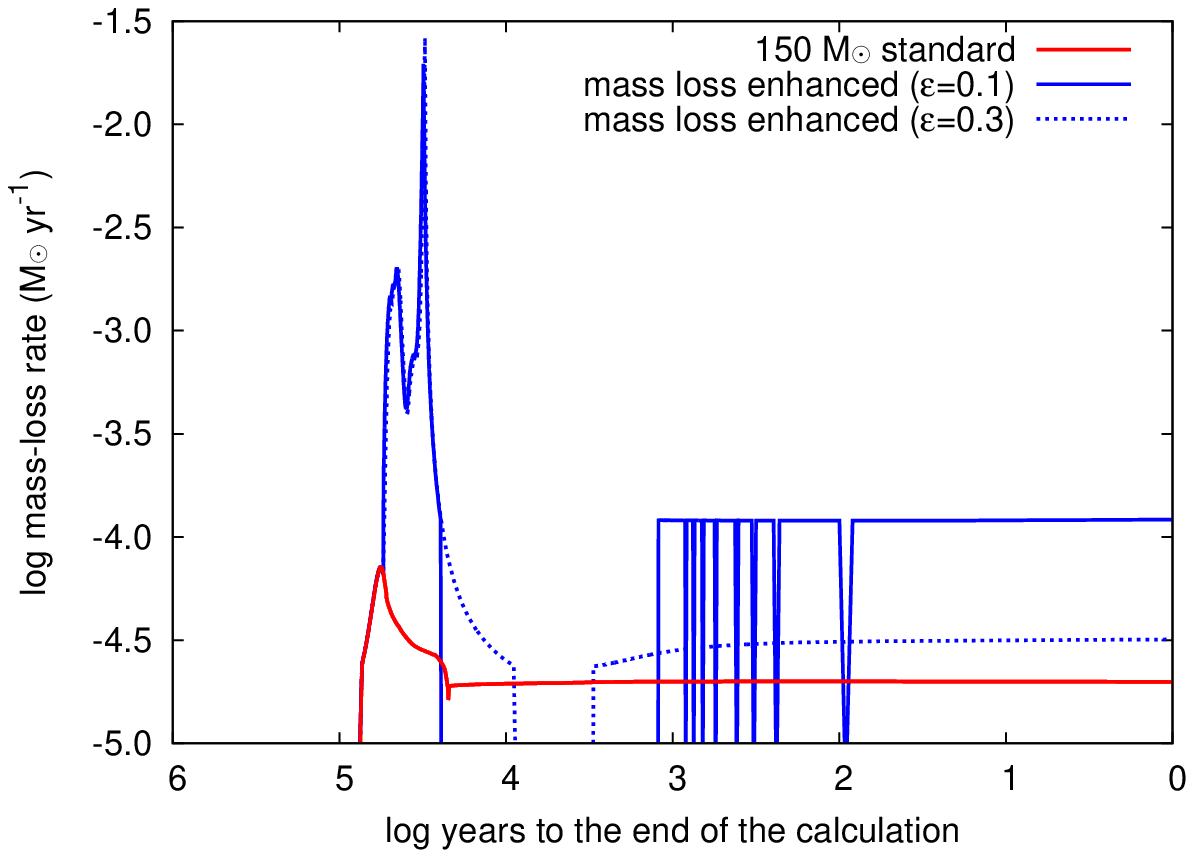}
\includegraphics[width=\columnwidth]{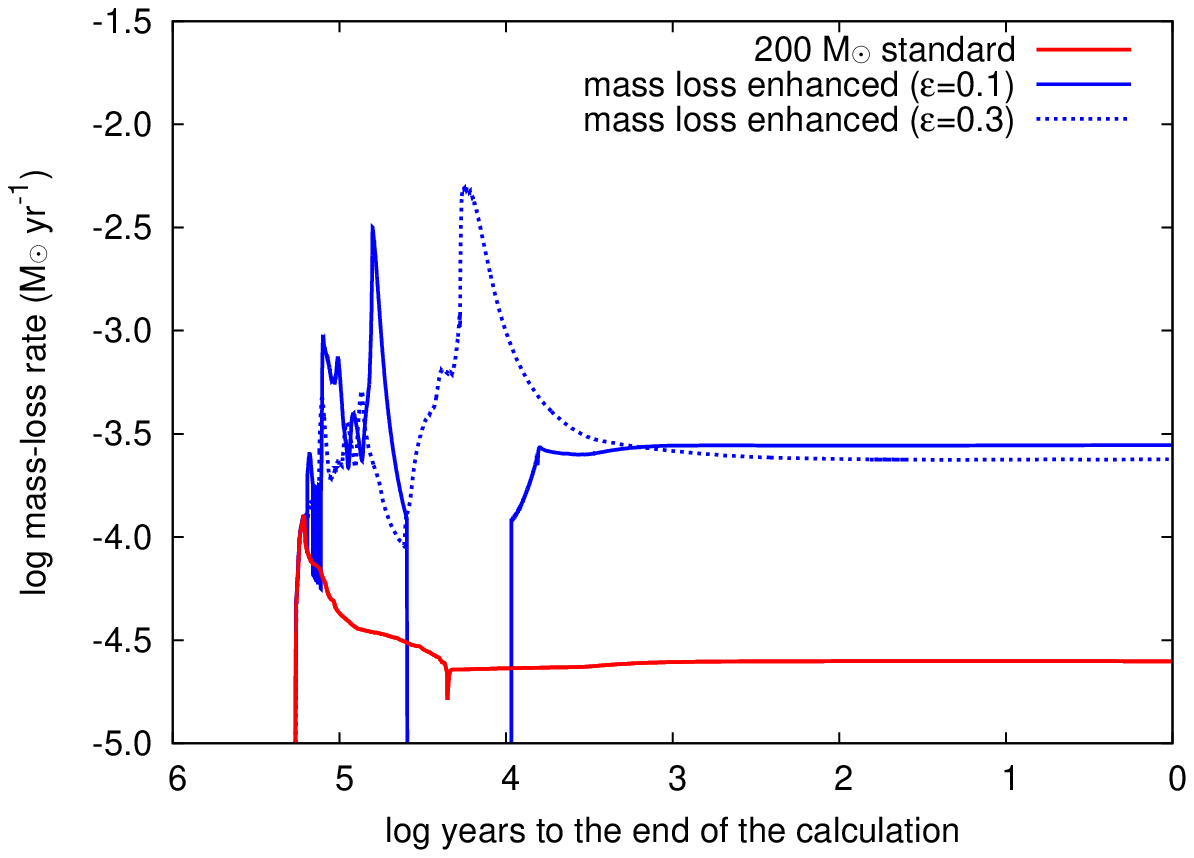}
\includegraphics[width=\columnwidth]{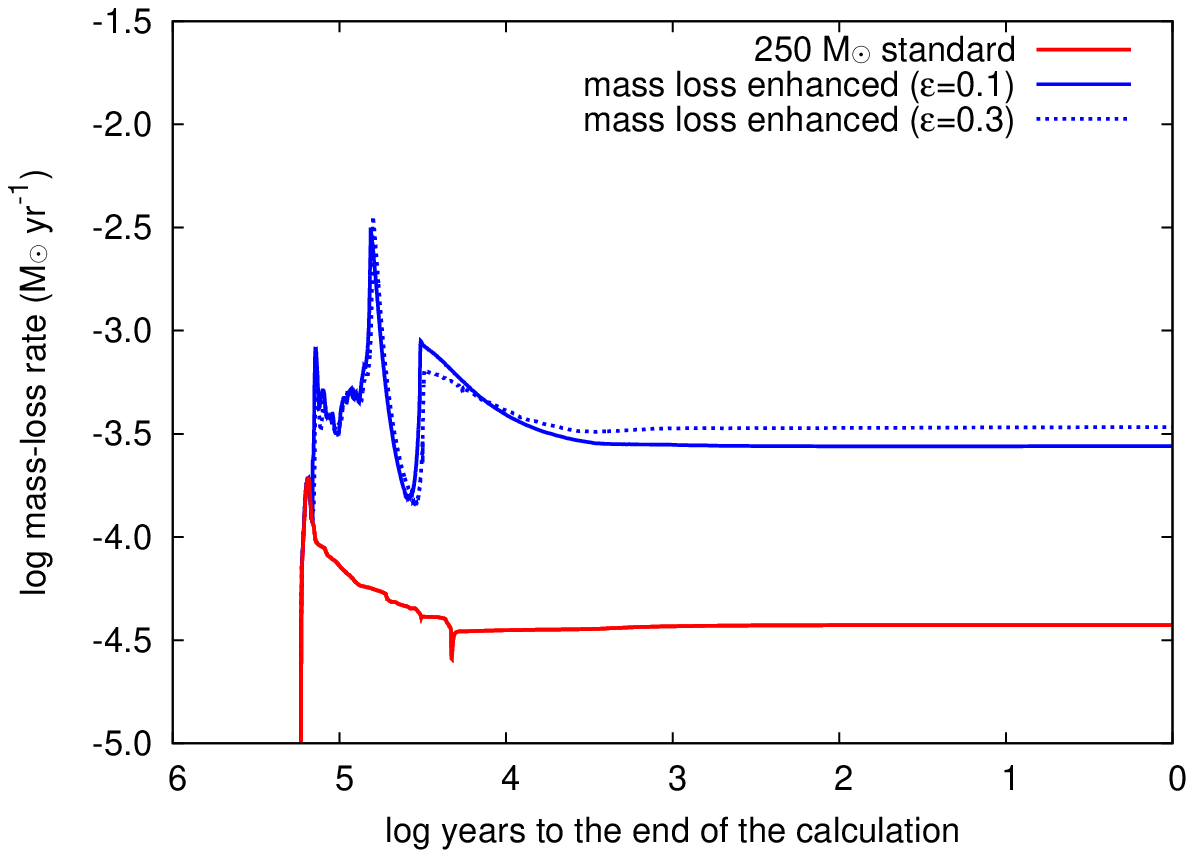}
\caption{
Comparisons of the mass-loss rates between the models with and without
the mass-loss enhancement. The mass-loss rates of Eq.~(\ref{eqenhancedmasslossrate})
with the indicated $\varepsilon$ are added in the enhanced models.
}
\label{enhancedmasslossrate}
\end{figure}

The results of these calculations are summarized in Table~\ref{tablestandard}.
We find that the stellar masses at oxygen core collapse are significantly reduced
compared to the models without pulsationally induced mass loss. 
The mass-loss histories of these models are presented in
Fig.~\ref{enhancedmasslossrate}.
The Kippenhahn diagrams for the models with $\varepsilon=0.1$ are
compared with those without mass-loss enhancement in Fig.~\ref{kippen}.
The mass-loss rates start to be
enhanced when hydrogen-rich envelopes become convective and
stars enter the RSG stage, as shown in the Kippenhahn diagrams.
During the final $10^{3}-10^{4}$ years, they are factors of 
$\sim 10$ larger than the radiation driven mass-loss rates, and amount
to $\sim 3\times 10^{-4}$ \mrunit.

The large differences in the mass-loss rates may cause
drastic changes in the circumstellar environment of PISN progenitors.
In most models with the mass-loss enhancement,
the CSM density becomes more than one order of magnitude higher.
The higher density
CSM will affect the observational properties of PISNe,
as we discuss in the next section.

The high pulsational mass-loss rates of PISN progenitors do not prevent
them from exploding as PISNe.
This is because the mass-loss enhancement is caused by the pulsations of
the hydrogen-rich envelope after the core hydrogen burning
and the mass-loss enhancement will terminate before
the entire hydrogen-rich envelope is ejected.
At this stage, the stars are not predicted to be pulsationally unstable.
Hence, the helium core mass
is not reduced as can be clearly seen in the Kippenhahn diagrams in Fig.~\ref{kippen}.
The pulsational mass loss only results in the
significant reduction in the hydrogen-rich envelope mass in PISN progenitors.

Although we focus on the PISN progenitor mass range in this paper, the 
pulsational amplitude growth is not limited to this mass range.
For example, \citet{yoon2012} show that the non-rotating
zero-metallicity stars with initial masses higher than 250 \Msun\
also evolve into RSGs. \citet{baraffe2001} also found that those higher
mass stars evolve to RSGs and become pulsationally unstable.
Thus, very massive stars can experience large mass loss as RSGs
even if they are metal-free.
The enhanced mass loss in these very massive stars may lead to
dust production in the early Universe even if they do not explode
eventually \citep{nozawa2014}.

\subsection{Metal-free pair-instability supernovae}
The high mass-loss rates in PISN progenitors are expected to have
significant influence on the observational properties of PISNe.
First, the extra mass loss significantly reduces the ejecta mass of PISNe.
The hydrogen-rich envelopes of $40-70$ \Msun\ are reduced in our
models with the mass-loss enhancement compared to those without.
The mass of the hydrogen-rich envelopes affect the light-curve
properties of PISNe like the duration of the plateau phase \citep[e.g.,][]{kasen2011}.

The mass which is ejected from the PISN progenitors forms their CSM.
They can sustain their high mass-loss rates
($\gtrsim 10^{-4}$ \mrunit\ with $\sim 100$ \kmps)
until the time of the explosion (Fig.~\ref{enhancedmasslossrate}).
Thus, they are embedded into a dense CSM when they explode.
The CSM density can be similar to that estimated in Type~IIn SNe
\citep[e.g.,][]{kiewe2012,taddia2013,moriya2014}
and the PISN observational properties are likely to be strongly affected
by the dense CSM.
For example, our 150~\Msun\ PISN progenitor has a 74~\Msun\ helium core.
It is expected to explode as a PISN, but a small amount of \Ni\
($\sim 0.1$~\Msun) is expected to be produced by the explosion
\citep{heger2002}. Thus, the SN does not get bright by
\Ni\ heating and it can only be as bright as $\simeq -18$ mag
in the optical due to shock heating \citep[e.g.,][]{kasen2011,kozyreva2014}.
However, if they have a dense CSM, the interaction between the SN ejecta and
the dense CSM can provide an extra heating source to make them brighter.
Since the explosion energy of PISNe
is very high ($\sim 10^{52}$ erg, \citealt{heger2002}),
the very energetic SN ejecta clashing into the dense CSM can actually make
even \Ni-poor PISN very bright.
The corresponding PISNe can be observed as bright Type~IIn SNe with
little \Ni.

The only suggested way to have dense CSM around the very
massive stars so far is related to the (core) pulsational pair instability \citep[e.g.,][]{woosley2007}.
This mechanism can work in massive stars which are slightly lighter than 
those which explode as PISNe, and they can lose a large amount of mass by
the central pair instability causing a partial ejection of the stellar material.
However, the stellar mass range to cause the pulsational pair instability is
limited \citep[e.g.,][]{heger2003,chatzopoulos2012}.
We find that higher mass stars can also lose a large
amount of mass due to (envelope) pulsations.
The progenitor mass range for very massive stars exploding within a dense
CSM is therefore wider than previously thought.

A critical difference between pulsational pair-instability SNe and
PISNe is whether they are actually accompanied by the explosions of the star.
Pulsational pair-instability SNe become bright due to the collision
of two or more massive shells ejected by the progenitor and they might
not be accompanied by actual SN explosions.
Conversely, PISNe are actual explosions which destroy the entire star 
without leaving any remnant. If the mass-loss rates of PISN
progenitors are enhanced by the pulsations, actual SN explosions occur
within a dense CSM.

High mass-loss rates of PISN progenitors may also end up with the
formation of a dense photoionization-confined shell if they are born
in an environment with a large ionizing-photon flux like massive star clusters
\citep{mackey2014}. Collision to such massive shell may result
in rebrightening of PISN light curves as discussed by \citet{mackey2014}.
The confined shells may also make PISNe very bright X-ray and/or radio transients
even if they are not dense enough to affect the optical luminosity
\citep[e.g.,][]{pan2013}.

Our finding that PISNe may have high mass-loss rates and
explode within a dense CSM even if they are
metal-free alters our expectation of the observational properties
of the first SNe in the Universe.
It has long been believed that the first SN progenitors do not suffer
much mass loss and the first SNe were considered to explode
in a sparse CSM. However, they may actually have a dense CSM environment
and not contain massive hydrogen-rich envelopes.
Especially, the existence of a dense CSM can make
PISNe brighter and bluer than previously expected and they may be easier to find
at very high redshifts than previously thought \citep[e.g.,][]{scannapieco2005,tanaka2012,tanaka2013,pan2012,hummel2012,desouza2013}.
\citet{cooke2009,cooke2012} may have already detected this kind of SNe up to $z=3.90$.
Many PISNe from the first stars
can be SLSNe with Type~IIn SN spectra and they can be easier to identify.
We investigate the detailed observational properties of PISNe exploding within
a dense CSM in a forthcoming paper.

\subsection{Superluminous supernovae and finite-metallicity pair-instability supernovae}
Type~IIn SLSNe like SN 2006gy are estimated to have both
dense CSM ($\sim 0.1$ \mrunit\ with 100 \kmps)
and very high explosion energy [$\sim (5-10)\times 10^{51}$ erg, 
e.g., \citealt{smith2007,chevalier2011,ginzburg2012,moriya2013,chatzopoulos2013,moriya2014b}].
Energetic explosions with little \Ni\ and a dense CSM are
qualitatively expected from our 150 \Msun\ models.
However, the maximum mass-loss rate
we obtained is $2.5\times 10^{-2}$ \mrunit\ (Fig.~\ref{eta-massloss}), which is still one order of
magnitude lower than those estimated for SLSNe of Type~IIn.
To obtain mass-loss rates as high as 0.1 \mrunit,
$\eta$ needs to be $0.24-0.28$ depending on $\varepsilon$ (Fig.~\ref{eta-massloss}).
The required $\eta$ and the $\eta$-\Teff\ relation (Eq.~\ref{eqeta}) indicate
that corresponding RSG PISN progenitors have lower effective temperature than $\Teff\simeq4600-4700$ K
to have the mass-loss rates higher than 0.1 \mrunit.
Although none of our models have effective temperature as low as
$4600-4700$ K because of the extremely low metallicity,
higher-metallicity RSG PISN progenitors can be cooler than
$4600-4700$ K \citep{langer2007} and they
are therefore likely to have mass-loss rates as high as 0.1 \mrunit.
Thus, they are a strong candidate for the progenitor of Type~IIn SLSNe.

Another interesting feature of observed SLSNe is that
those that have slowly declining light curves which are
consistent with large production of \Ni\ are all Type~Ic \citep[e.g.,][]{gal-yam2012}.
This indicates that PISNe with massive cores producing a large amount of
\Ni\ may not explode with their hydrogen-rich envelope.
Although none of our models lose all of their hydrogen-rich envelope,
the hydrogen-rich envelopes may be removed more easily with the help of
the larger radiation-driven mass-loss rates
in higher metallicity PISN models \citep{langer2007}.
Thus, the local PISNe may tend to have little or no hydrogen left when
they explode because of the pulsational mass-loss enhancement
with radiation-driven mass loss,
making them primarily Type~Ic.
If their cores are massive enough to have large \Ni, they are observed as
Type~Ic SLSNe with large \Ni\ mass. If PISN progenitors with low-mass cores
succeed in losing their hydrogen-rich envelopes, they do not produce
much \Ni\ but they can still be observed as relatively faint
Type~Ic SNe because of shock heating \citep{herzig1990}.

Some Type~Ic SLSNe have rapidly declining light curves which are
not consistent with \Ni\ heating \citep[e.g.,][]{quimby2011,pastorello2010,chomiuk2011}.
There are several mechanisms said
to explain their high luminosities without \Ni\ heating \citep[e.g.,][]{kasen2010,dessart2012,inserra2013,nicholl2013,metzger2014}.
One possibility is that SLSN progenitors are surrounded by a hydrogen-free
dense CSM \citep[e.g.,][]{blinnikov2010,leloudas2012,moriya2012,ginzburg2012,benetti2014}.
Pulsational pair-instability SNe from large hydrogen-free cores have
been argued to make such a hydrogen-free dense CSM
(e.g., \citealt{chatzopoulos2012b}, see also \citealt{chevalier2012}
for another suggested way).
Hydrogen-free cores massive enough to induce
pulsational instability SNe may be created by
rapidly-rotating stars (e.g., \citealt{yoon2012,chatzopoulos2012}, see
also \citealt{yoon2006}).
As discussed previously, we argue that hydrogen-free cores may also come from
non-rotating stars with the pulsation-driven mass-loss enhancement.
Very massive stars causing pulsational pair-instability SNe also evolve
to RSGs \citep[e.g.,][]{yoon2012,chatzopoulos2012}
and they can lose a significant amount of their hydrogen-rich
envelope during the RSG stage. The subsequent core pulsational instability
may end up with the creation of a massive hydrogen-free CSM 
which is said to account for rapidly declining Type~Ic SLSNe.

\section{Conclusions}\label{conclusions}
We have shown that PISN progenitors during the RSG stage are pulsationally
unstable. A growing pulsational amplitude may induce mass loss 
and even metal-free PISN progenitors can have high mass-loss rates.

We find that metal-free PISN progenitors can be pulsationally unstable
when their effective temperature
 is lower than $\simeq 5000$ K. The growth rate of the pulsations
is found to strongly correlate with \Teff\ (Eq.~\ref{eqeta}).
If part of the kinetic energy gain of the pulsation
is used to initiate mass loss, the mass-loss rate can be
very high even if stars are metal-free and the radiation-driven mass
loss is inefficient (Eq.~\ref{eqenhancedmasslossrate} and Fig.~\ref{eta-massloss}).

The mass-loss rates of our PISN progenitor models within $\sim 10^3$ years
before the explosions become higher than $10^{-4}$ \mrunit
when the pulsation-driven mass loss is taken into account (Fig.~\ref{enhancedmasslossrate}).
Because the mass-loss enhancement by the pulsations
is initiated in the hydrogen-rich envelope after the hydrogen burning,
the mass-loss enhancement stops before the entire
hydrogen-rich envelope is lost. Thus, the helium core mass is not
reduced by the pulsation-driven mass loss and
the stars can still explode as PISNe (Fig.~\ref{kippen}).

The pulsation-driven mass loss significantly reduces the
masses of PISN progenitors at the time of their explosions (Table~\ref{tablestandard}).
Moreover, it forms a dense CSM around PISN progenitors,
which can significantly affect the observational properties of the SNe.
The existence of a dense CSM around PISNe can make them brighter and
bluer, making them easier to observe at high redshifts than previously thought.
Our results also show that metal-free very massive stars do not need to be
within a limited mass range of pulsational pair-instability SNe to have a dense CSM.

Type~IIn SLSNe are estimated to have dense CSM ($\sim 0.1$ \mrunit with 100 \kmps),
high explosion energy [$\sim (5-10)\times 10^{51}$ erg], and little \Ni\ production.
If we account for the pulsation-driven mass loss, these features are
expected from low-mass PISN progenitors around 150 \Msun.
Although our metal-free PISN progenitors do not become cool enough to
make the mass-loss rate as high as $\sim 0.1$ \mrunit, the higher-metallicity PISN
progenitors are likely to be cool enough to obtain such high mass-loss rates.
PISN candidates with a large \Ni\ production found
so far are all Type~Ic SNe and they do not have hydrogen at all. They may
not have hydrogen
because the hydrogen-rich envelope is taken away by the pulsation-driven mass loss.
Pulsational pair-instability SNe in such hydrogen-free cores may 
end up with rapidly declining Type~Ic SLSNe.
Detailed properties of PISNe within a dense CSM from pulsation-driven mass enhancement
will be presented in our forthcoming paper.

\begin{acknowledgements}
We would like to thank the referee, Andr\'e Maeder, for constructive comments.
TJM is supported by Japan Society for the Promotion of
 Science Postdoctoral Fellowships for Research Abroad
 (26\textperiodcentered 51).
Numerical computations were carried out on computers at Center for Computational Astrophysics, National Astronomical Observatory of Japan.
\end{acknowledgements}

\Online

\begin{appendix} 
\section{Input parameters for \texttt{MESAstar}}\label{appendix}
An example of the input parameters for \texttt{MESAstar} used
in this study is shown. Note that the code needs to be modified to
incorporate, e.g., the metallicity dependence of the radiation-driven mass-loss
rates below $\Teff=10^4$ K.

{\small
\begin{verbatim}
&star_job

         create_pre_main_sequence_model = .true.

         change_v_flag = .true.
         new_v_flag = .true.

         change_net = .true.
         new_net_name = 'approx21.net'

         kappa_file_prefix = 'gs98'
         
/ ! end of star_job namelist

&controls

         initial_mass = 150
         initial_z = 0.00

         RGB_wind_scheme = 'Dutch'
         AGB_wind_scheme = 'Dutch'
         Dutch_wind_lowT_scheme = 'de Jager'
         RGB_to_AGB_wind_switch = 1d-4
         Dutch_wind_eta = 1.0 
         de_Jager_wind_eta = 1.0 

         MLT_option = 'Henyey'
         mixing_length_alpha = 1.6
         use_Ledoux_criterion = .false.
         alpha_semiconvection = 0.0 

         use_Type2_opacities = .true.
         Zbase = 0.00

         which_atm_option = 'Eddington_grey'               

         max_years_for_timestep = 1d4 !1d-3
         min_timestep_limit = 1d-12 
         min_timestep_factor = 0.7d0
         max_timestep_factor = 1.2d0

         !use_artificial_viscosity = .true.
         !l1_coef = 0.
         !l2_coef = 2.

         varcontrol_target = 1d-4
         dX_nuc_drop_limit = 1d-2

         he_core_boundary_h1_fraction = 1d-4

         mesh_delta_coeff = 1.00

         mesh_dlog_pp_dlogP_extra = 0.21  
         mesh_dlog_cno_dlogP_extra = 0.21 
         mesh_dlog_3alf_dlogP_extra = 0.21 
         mesh_dlog_burn_c_dlogP_extra = 0.21 
         mesh_dlog_burn_n_dlogP_extra = 0.21 
         mesh_dlog_burn_o_dlogP_extra = 0.21 
         mesh_dlog_burn_ne_dlogP_extra = 0.21 
         mesh_dlog_burn_na_dlogP_extra = 0.21 
         mesh_dlog_burn_mg_dlogP_extra = 0.21 
         mesh_dlog_cc_dlogP_extra = 0.21 
         mesh_dlog_co_dlogP_extra = 0.21 
         mesh_dlog_oo_dlogP_extra = 0.21 
         mesh_dlog_burn_si_dlogP_extra = 0.21 
         mesh_dlog_burn_s_dlogP_extra  = 0.21 
         mesh_dlog_burn_ar_dlogP_extra = 0.21 
         mesh_dlog_burn_ca_dlogP_extra = 0.21 
         mesh_dlog_burn_ti_dlogP_extra = 0.21 
         mesh_dlog_burn_cr_dlogP_extra = 0.21 
         mesh_dlog_burn_fe_dlogP_extra = 0.21 
         mesh_dlog_pnhe4_dlogP_extra = 0.21 
         mesh_dlog_other_dlogP_extra = 0.21 
         mesh_dlog_photo_dlogP_extra = 0.21 

/ ! end of controls namelist
\end{verbatim}
}

\end{appendix}

\end{document}